\documentstyle[emulateapj,psfig,natbib,amsfonts,amsmath]{article}
\def\ra#1#2#3{#1$^{\rm h}$#2$^{\rm m}$#3$^{\rm s}$}
\def\dec#1#2#3{$#1^\circ#2'#3''$}
\def\swift{{\it Swift}}
\def\nod{\nodata}
\def\grb{GRB\,051111}

\def\pom{1}
\def\ociw{2}
\def\prince{3}
\def\hubble{4}
\def\psu{5}
\def\cit{6}
\def\keck{7}

\begin{document}

\title{Spectroscopy of GRB\,051111 at $z=1.54948$: Kinematics and 
Elemental Abundances of the GRB Environment and Host Galaxy}

\author{
B.~E.~Penprase\altaffilmark{\pom},
E.~Berger\altaffilmark{\ociw,}\altaffilmark{\prince,}\altaffilmark{\hubble},
D.~B.~Fox\altaffilmark{\psu},
S.~R.~Kulkarni\altaffilmark{\cit},
S. Kadish\altaffilmark{\pom},
L. Kerber\altaffilmark{\pom},
E.~Ofek\altaffilmark{\cit},
M.~Kasliwal\altaffilmark{\cit},
G.~Hill\altaffilmark{\keck},
B. Schaefer\altaffilmark{\keck},
and M. Reed\altaffilmark{\keck}
}

\altaffiltext{\pom}{Pomona College Department of Physics and Astronomy,
610 N. College Avenue, Claremont, CA}

\altaffiltext{\ociw}{Observatories of the Carnegie Institution
of Washington, 813 Santa Barbara Street, Pasadena, CA 91101}
 
\altaffiltext{\prince}{Princeton University Observatory,
Peyton Hall, Ivy Lane, Princeton, NJ 08544}
 
\altaffiltext{\hubble}{Hubble Fellow}

\altaffiltext{\psu}{Department of Astronomy and Astrophysics,
Pennsylvania State University, 525 Davey Laboratory, University
Park, PA 16802}

\altaffiltext{\cit}{Division of Physics, Mathematics and Astronomy,
105-24, California Institute of Technology, Pasadena, CA 91125}

\altaffiltext{\keck}{W.~M.~Keck Observatory, 65-1120 Mamalahoa
Highway, Kamuela, HI 96743}

\begin{abstract}
We present a high-resolution, high signal-to-noise optical spectrum of
the afterglow of \grb\ obtained with the HIRES spectrograph on the
Keck I 10-m telescope.  The spectrum exhibits three redshifted
absorption systems with the highest, at $z=1.54948$, arising in the
GRB host galaxy.  While the Ly$\alpha$ feature is outside the range of
our spectrum, the high column density of weakly-depleted Zn suggests
that the host is a damped Ly$\alpha$ system with $N({\rm HI})\gtrsim
10^{21}(Z/Z_\odot)^{-1}$.  The bulk of the gas ($>80\%$) is confined
to a narrow velocity range of $|v|<30$ km s$^{-1}$ exhibiting strong
dust depletion of refractory elements such as Fe and Cr.  The
depletion pattern is similar to that observed in warm disk clouds of
the Milky Way.  We also detect absorption from all ground-level
fine-structure states of \ion{Fe}{2}, the first such example in a
QSO-DLA or GRB-absorption spectrum, which indicate conditions that are
consistent with the "warm disk" depletion pattern.  The absorption
profiles of \ion{Fe}{2} and \ion{Mg}{2} extend over several hundred km
s$^{-1}$, with a depletion pattern that more closely resembles that of
QSO-DLAs, suggesting that the sight line to \grb\ probes the halo of
the host galaxy in addition to the dense disk.  Thus, detailed
diagnostics of the interstellar medium of GRB host galaxies continue
to provide insight into regions which are generally missed in quasar
surveys.
\end{abstract}
 

\section{Introduction}
\label{sec:intro}

Most of the data available for elemental abundances in the early
universe come from studies of damped Ly$\alpha$ (DLA) systems detected
in absorption against background quasars \citep{wgp05}.  DLAs appear
to be metal poor, with a typical $Z\sim 0.03$ $Z_\odot$
\citep{pgw+03}, and while they exhibit evidence for star formation in
a few cases, the rates typically appear to be lower than in the Lyman
break galaxies at a similar redshift range \citep{bwc+99}.  Since the
existence of a DLA system depends on a chance alignment of the quasar
and DLA gas, and this is affected by a cross-section bias and possibly
dust, it is unclear at present whether the QSO-DLAs are a
representative sample of the interstellar medium within early
galaxies.

An alternative approach to probing intervening gas in galaxies and the
IGM is to use the afterglows of gamma-ray bursts (GRBs).  In the
context of the relation to star formation and the nature of DLAs, GRBs
offer several advantages over quasar studies.  First, GRBs are
embedded in star forming galaxies with typical offsets of a few kpc or
less \citep{bkd02}.  They therefore not only provide a direct link to
star formation, but also probe the regions of most intense star
formation, and hence the production and dispersal of metals.  Second,
since the GRB afterglow emission fades away on a timescale of days to
weeks, the host galaxy and any intervening DLAs can be subsequently
studied directly (e.g., \citealt{vel+04}).

Third, and perhaps most important, since GRBs are likely to be located
in star forming regions within their host galaxies, this approach
provides the only systematic way to directly probe the small-scale
environment and conditions of star formation at high redshift; the
probability of intersecting an individual star-forming cloud in a
quasar sight line is vanishingly small.

Over the past several years, a few absorption spectra of GRB
afterglows have been obtained, revealing relatively large neutral
hydrogen column densities, in some cases with ${\rm log}\,N({\rm
HI})>22$ \citep{vel+04,bpc+05,cpb+05}.  The metallicity, inferred in
only a few cases, appears to be sub-solar ($Z\sim 0.01-0.1$ $Z_\odot$;
\citealt{vel+04,bpc+05,cpb+05,sve+05}), but with a dust-to-gas ratio 
that is larger than that in QSO-DLAs \citep{sff03}.  In addition, some
spectra reveal complex velocity structure, interpreted to arise from
ordered galactic rotation \citep{cgh+03}, and in some cases appearing
to arise in the complex wind environment of the progenitor star
\citep{mfh+02,bpc+05}.  

In a continuing effort to characterize the interstellar medium of high
redshift galaxies and to provide a comparison to QSO-DLAs, we present
here a Keck HIRES absorption spectrum of \grb, which reveals
absorption from a possible DLA with a column density ${\rm
log}\,N({\rm HI})\gtrsim 21$ at a redshift of $z=1.54948$.  The
spectrum of \grb\ is remarkable not only for its high resolution,
which enables accurate estimates of abundances within the host galaxy,
but also for the detection and accurate measurement of excited
fine-structure \ion{Fe}{2} states \citep{bpf+05} and a detailed
examination of the kinematics within the interstellar medium of the
host galaxy.

\section{Observations}
\label{sec:obs}

\grb\ was detected by \swift\ on 2005 November 11 at 06:00:02 UT.  The
duration of the burst is 47 s, and the fluence in the 15-150 keV band
is $(3.9 \pm 0.1)\times 10^{-6}$ erg cm$^{-2}$ \citep{gcn4260}.
Observations with the ROTSE-IIIb robotic telescope started 27 seconds
after the burst, and revealed an uncatalogued fading source located at
$\alpha$=\ra{23}{12}{33.2}, $\delta$=\dec{+18}{22}{29.1} (J2000)
\citep{gcn4247}. 

Spectroscopic observations of \grb\ were initiated at 07:03 UT,
approximately one hour after the burst, using the High Resolution
Echelle Spectrometer (HIRES) mounted on the Keck I 10-m telescope
\citep{gcn4255}.  A total of 5400 s of exposure time were obtained
using a $0.86''$ wide slit at airmass $1.0-1.2$.  The wavelength range
covered is $4200-8400$ \AA.  The spectra were reduced using the MAKEE
pipeline routines (Version 4.0.1 of May 2005), which includes optimal
extraction of orders, sky subtraction, and wavelength calibration from
Th-Ar arc lamp exposures, including a heliocentric velocity
correction.  The orders within individual frames were traced using a
median combined total of the exposures, and atmospheric absorption
features were removed with the Makee pipeline.  A final resampling of
the spectrum and continuum fitting was performed using the IRAF task
{\tt continuum}.  The column densities of species detected in the
spectrum were computed using a set of custom-written IDL routines for
measuring equivalent widths, optical depths, and for fitting curve of
growth models to the detected absorption species.

\begin{figure}
%\epsscale{1.0}
%\plotone{figbep1.ps}
\centerline{\psfig{file=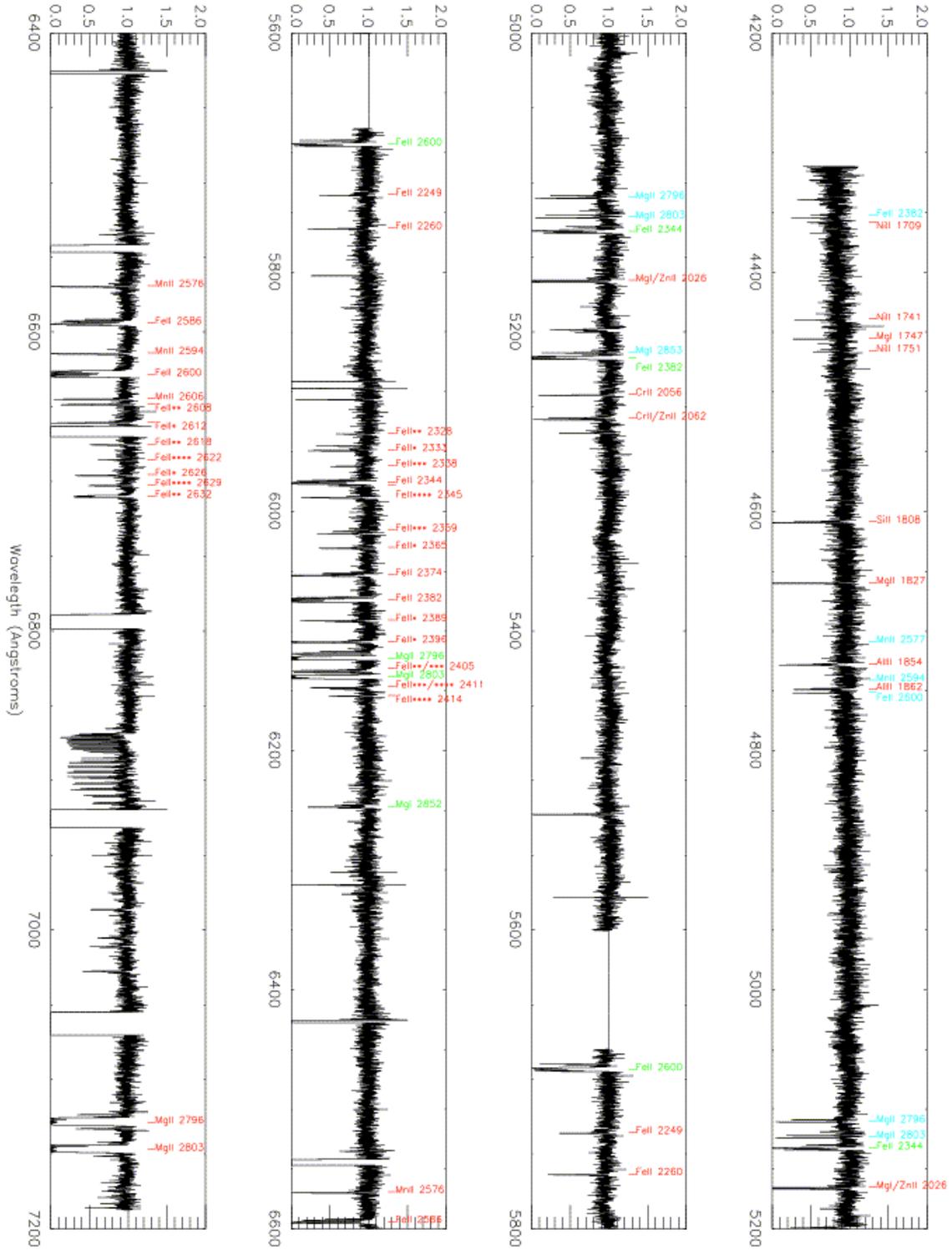,width=6.0in}}
\caption{Complete spectrum of \grb\ obtained with HIRES on the Keck I 
telescope showing the spectrum of chips 1 and 2, which spans
wavelengths from 4300 to 7200 \AA. Absorption lines from the system at
$z_1=1.54958$ are indicated in red, while features from the system at
$z_2=1.18975$ are indicated in green and those from $z_{3ab}=0.827$
are shown in blue.
\label{fig:spec}}
\end{figure}

The spectrum, shown in Figure~\ref{fig:spec}, reveals strong
absorption features at a redshift, $z_1=1.54948\pm 0.00001$, which we
interpret as the redshift of the host galaxy (see also
\citealt{gcn4271}\footnotemark\footnotetext{A GRB Coordinates Network 
(GCN) Circular by \citet{gcn4271} provides a cursory list of some
properties of the host galaxy absorption system.}).  In addition, we
detect intervening systems at $z_2=1.18975$ (\ion{Mg}{2},
\ion{Mg}{1}, and \ion{Fe}{2}), $z_{3a}=0.82761$ (\ion{Mg}{1},
\ion{Mg}{2}, and \ion{Fe}{2}) and $z_{3b}=0.82698$ (\ion{Mg}{1} and
\ion{Mg}{2}).  The latter two most likely arise from the same object
given their close redshifts.  In addition to the GRB absorption
spectrum we also obtained imaging of the GRB field with the Echellete
Spectrograph and Imager on the Keck II telescope to search for host
galaxy emission as discussed in \S\ref{sec:host}.

\section{Kinematics of the Host Galaxy Absorption Line System}
\label{sec:kin}

The strongest absorption features in the spectrum of \grb\ arise from
the redshift system $z_1=1.54948$, which exhibits a wide array of
metal lines of \ion{Mg}{1}, \ion{Mg}{2}, \ion{Mn}{2}, \ion{Cr}{2},
\ion{Fe}{2}, \ion{Zn}{2}, \ion{Al}{3}, \ion{Si}{2}, \ion{Ni}{2}, and 
the four ground-level fine-structure states of \ion{Fe}{2} and one of
\ion{Si}{2} \citep{bpf+05}.  The line identifications, observed 
wavelengths, and equivalent widths (EWs) are listed in
Table~\ref{tab:lines}.  Also included in Table~\ref{tab:lines} are
estimates of the column densities based on the weak line limit for
each transition using oscillator strengths and rest wavelengths from
\citet{mor91}.

\begin{figure}
%\epsscale{0.9}
%\plotone{figbep2a.ps}
\centerline{\psfig{file=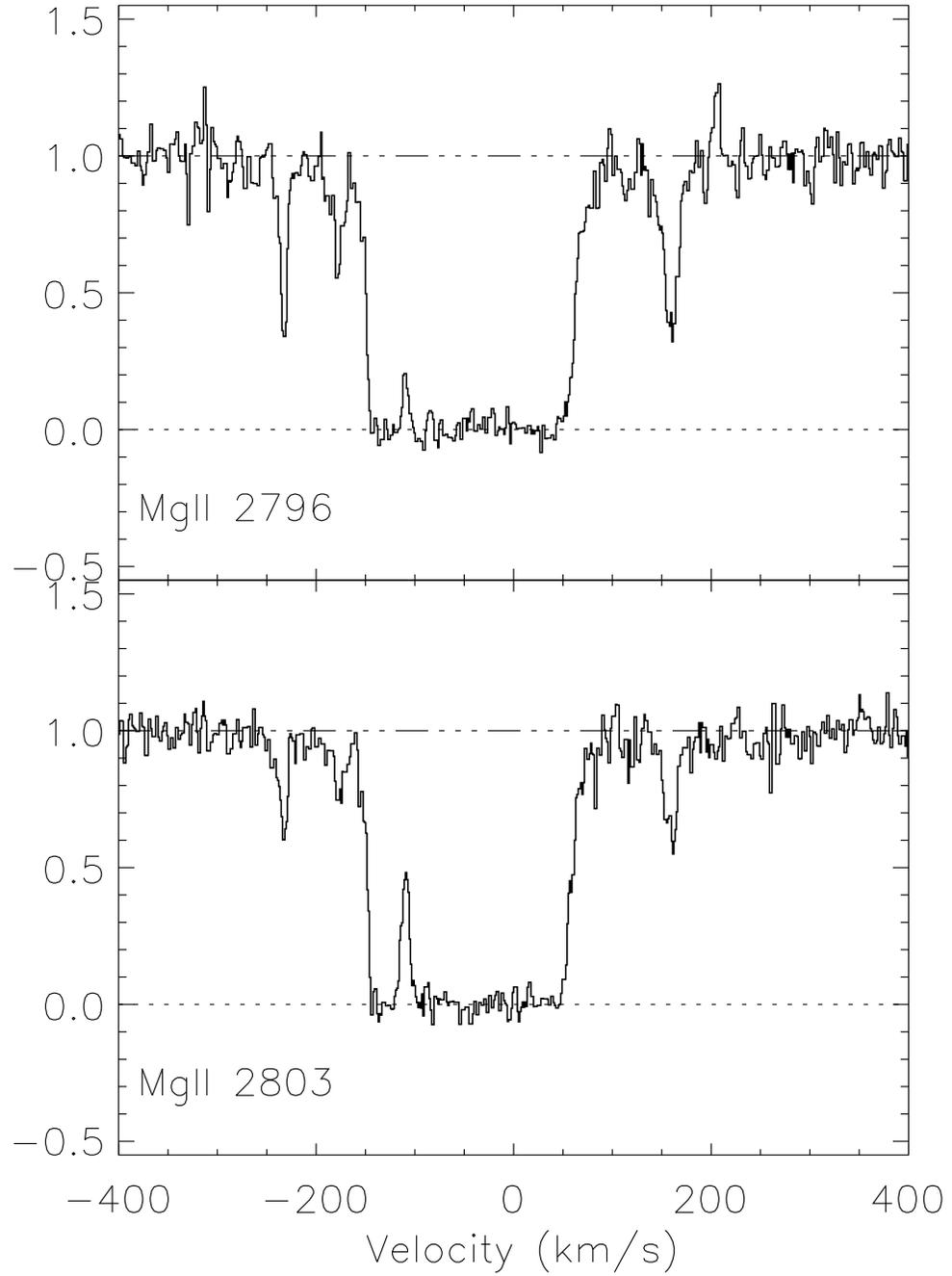,width=6.0in}}
\caption{Spectrum of \ion{Mg}{2} absorption within host galaxy of \grb\ 
in the rest frame of $z_1=1.54948$.  Observational details are given in 
\S\ref{sec:obs}.  
\label{fig:MgIIspec}}
\end{figure}

\begin{figure}
%\epsscale{0.9}
%\plotone{figbep2b.ps}
\centerline{\psfig{file=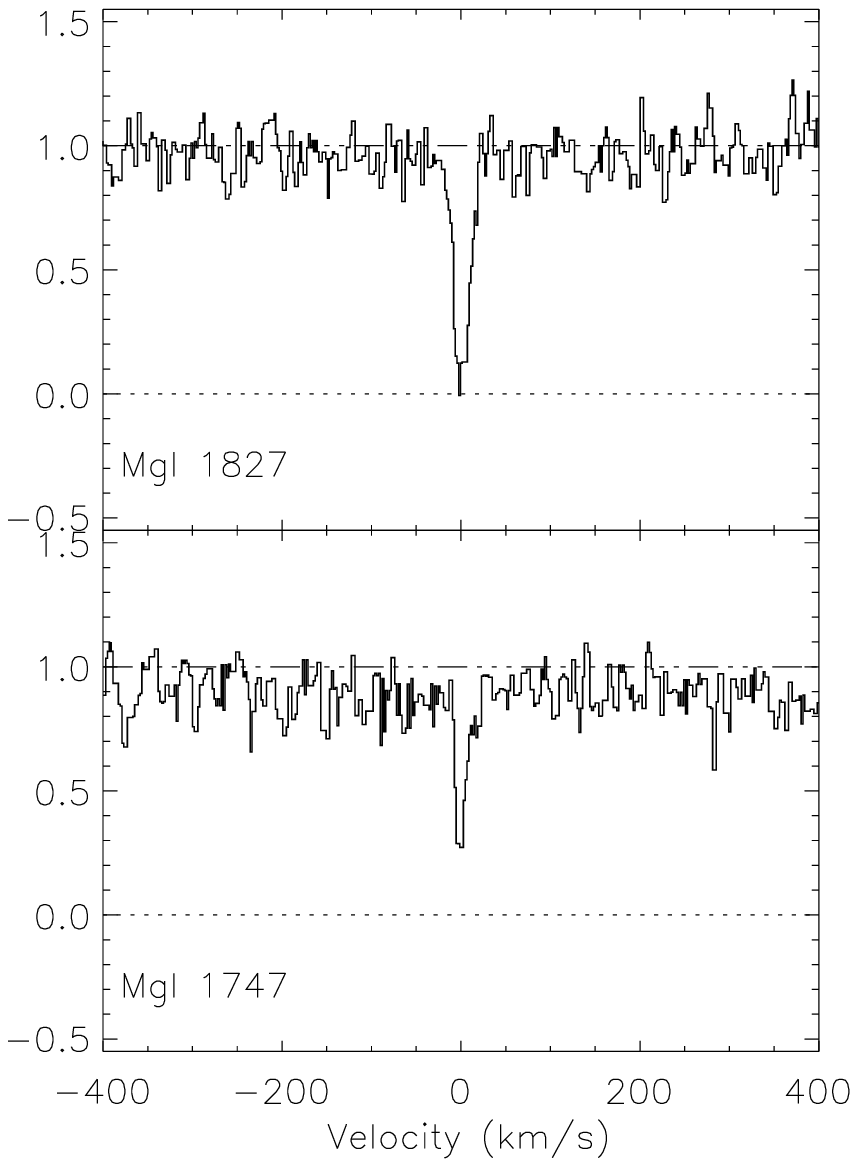,width=6.0in}}
\caption{Same as Figure~\ref{fig:MgIIspec} but for \ion{Mg}{1} 
transitions. 
\label{fig:MgIspec}}
\end{figure}

\begin{figure}
%\epsscale{0.9}
%\plotone{figbep2c.ps}
\centerline{\psfig{file=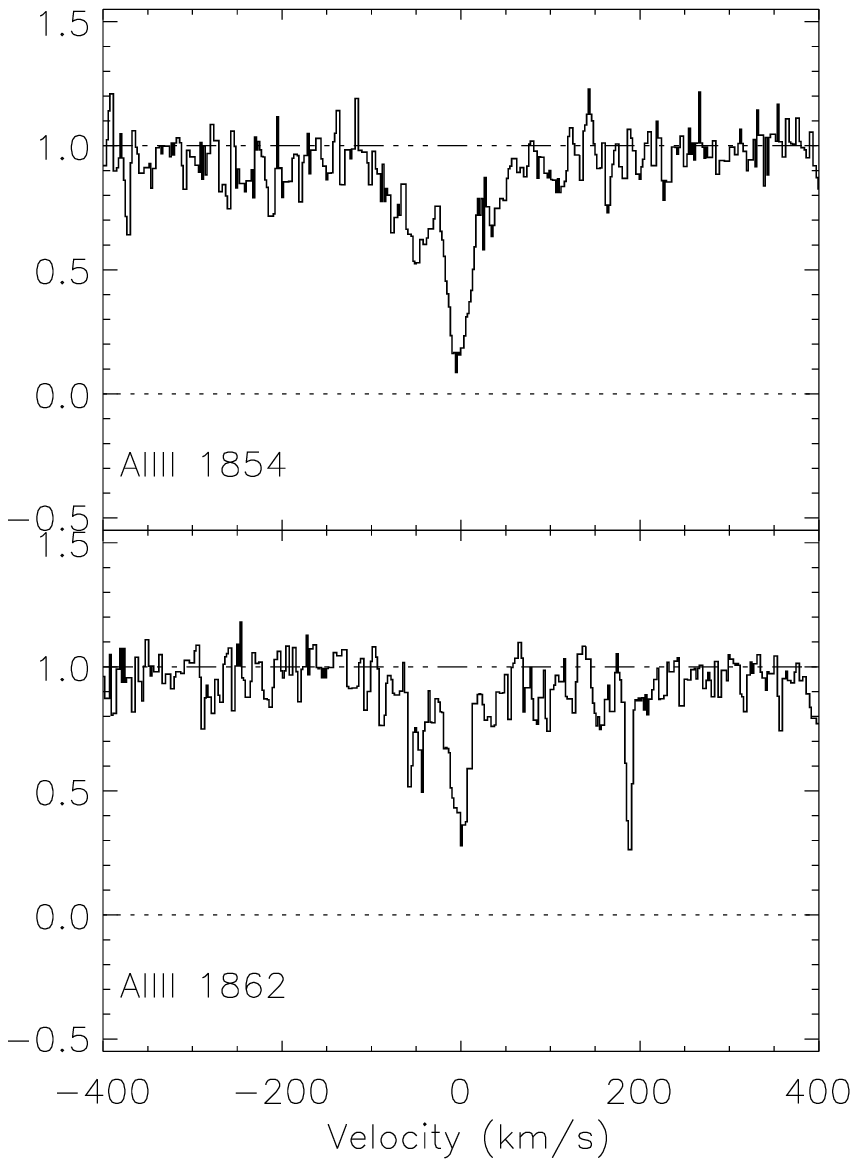,width=6.0in}}
\caption{Same as Figure~\ref{fig:MgIIspec} but for \ion{Al}{3} 
transitions. 
\label{fig:AlIIIspec}}
\end{figure}

\begin{figure}
%\epsscale{0.9}
%\plotone{figbep2d.ps}
\centerline{\psfig{file=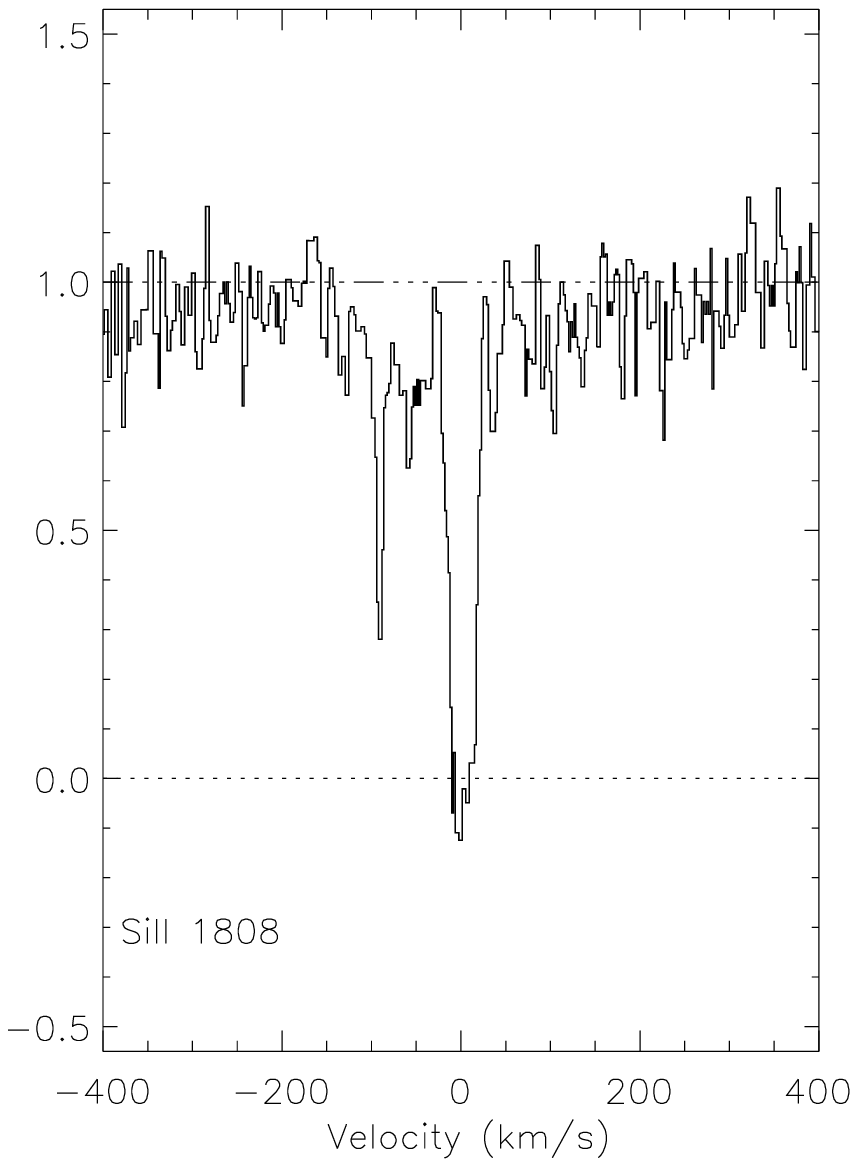,width=6.0in}}
\caption{Same as Figure~\ref{fig:MgIIspec} but for \ion{Si}{2} 
transitions. 
\label{fig:Sispec}}
\end{figure}

\begin{figure}
%\epsscale{0.9}
%\plotone{figbep2e.ps}
\centerline{\psfig{file=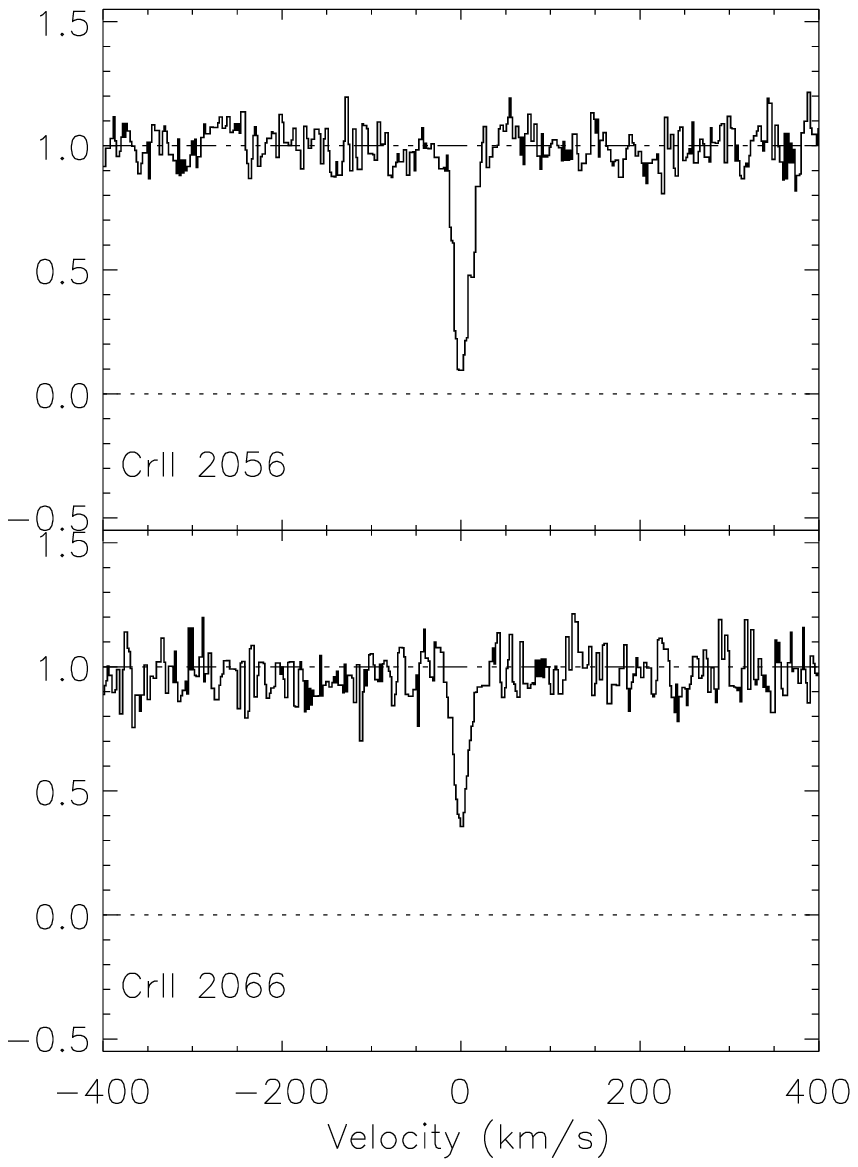,width=6.0in}}
\caption{Same as Figure~\ref{fig:MgIIspec} but for \ion{Cr}{2} 
transitions. 
\label{fig:CrIIspec}}
\end{figure}

\begin{figure}
%\epsscale{0.9}
%\plotone{figbep2f.ps}
\centerline{\psfig{file=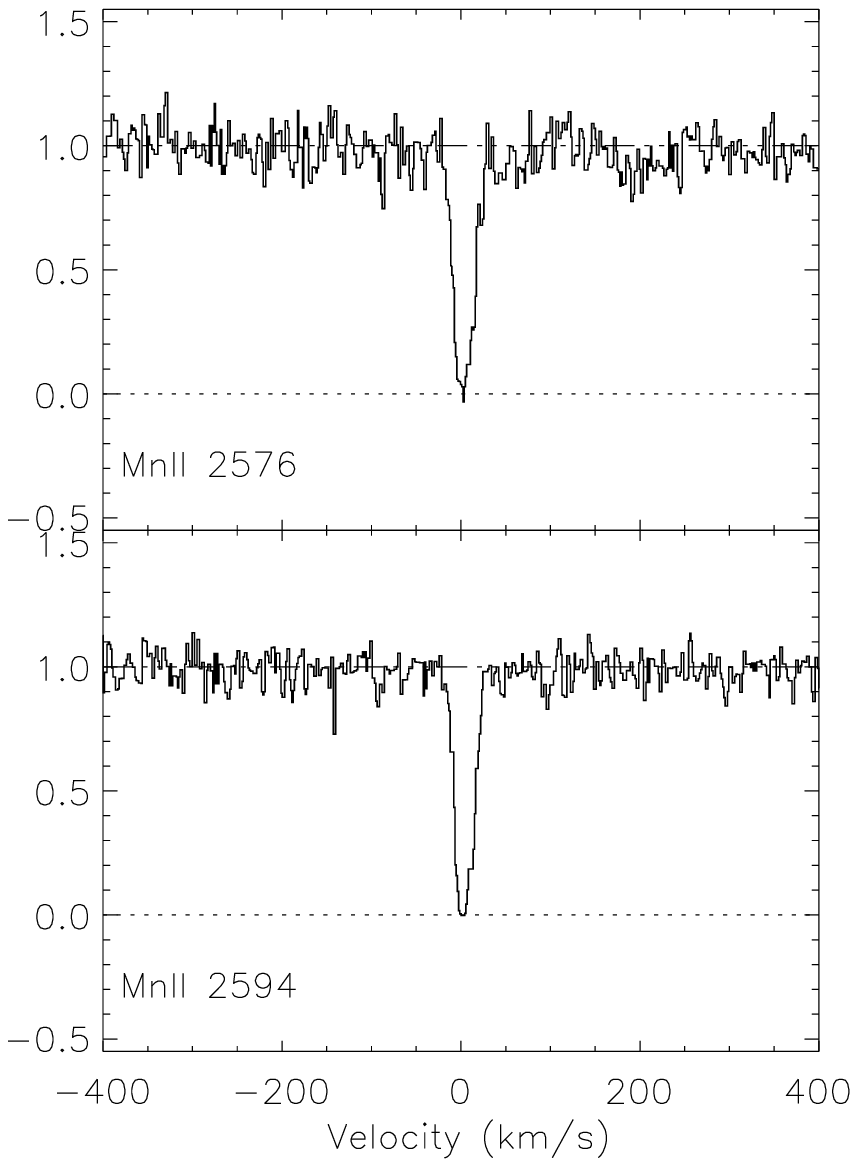,width=6.0in}}
\caption{Same as Figure~\ref{fig:MgIIspec} but for \ion{Mn}{2} 
transitions.
\label{fig:MnIIspec}}
\end{figure}

\begin{figure}
%\epsscale{0.9}
%\plotone{figbep2g.ps}
\centerline{\psfig{file=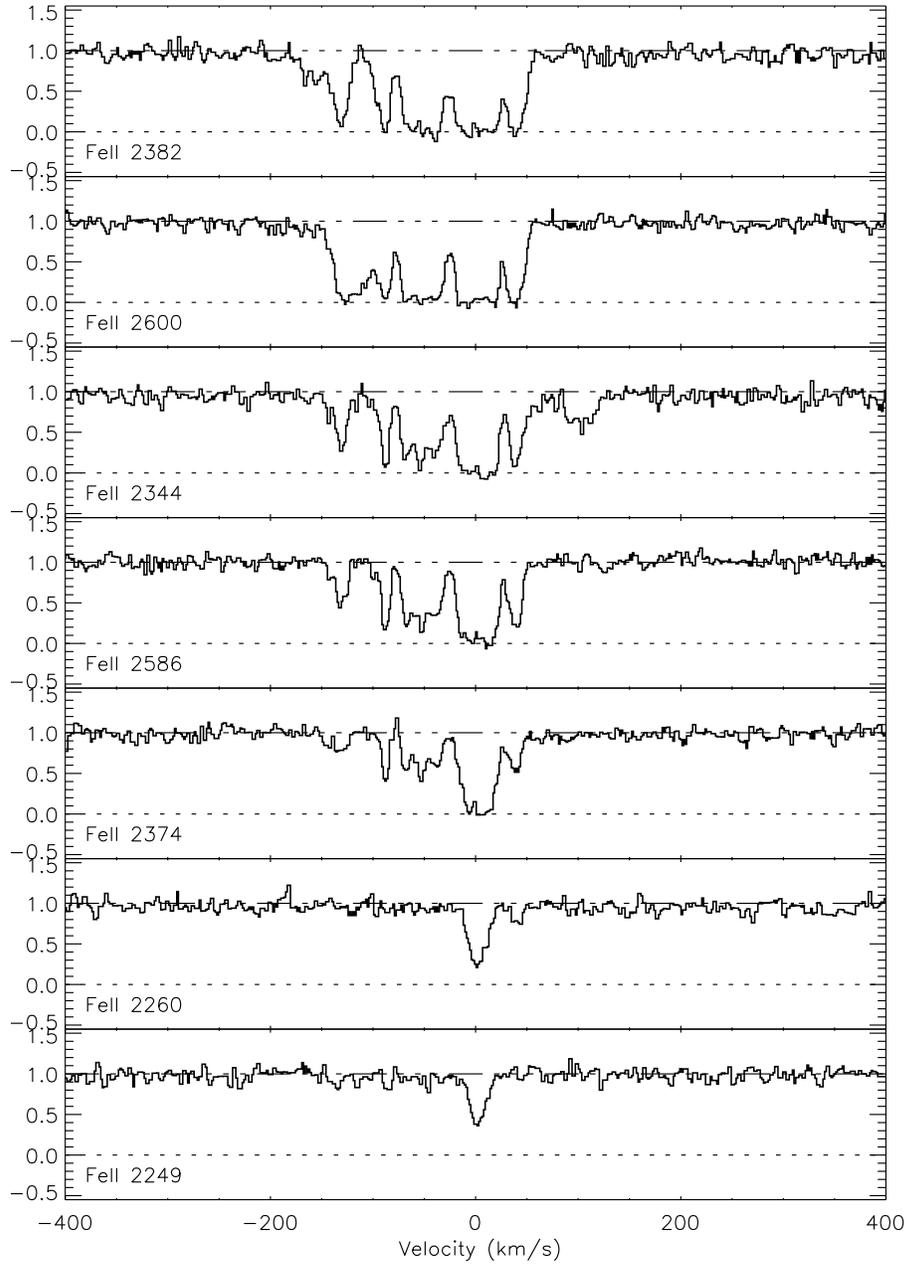,width=6.0in}}
\caption{Same as Figure~\ref{fig:MgIIspec} but for \ion{Fe}{2} 
transitions.  The \ion{Fe}{2} absorption has a velocity structure
similar to that of \ion{Mg}{2}, and the large number of transitions of
differing oscillator strength enables recovery of the kinematics in
the dense core from $-30<v<30$ km s$^{-1}$, as well as the outlying
absorption from $-200<v<60$ km s$^{-1}$.
\label{fig:FeIIspec}}
\end{figure}

\begin{figure}
%\epsscale{0.9}
%\plotone{figbep2h.ps}
\centerline{\psfig{file=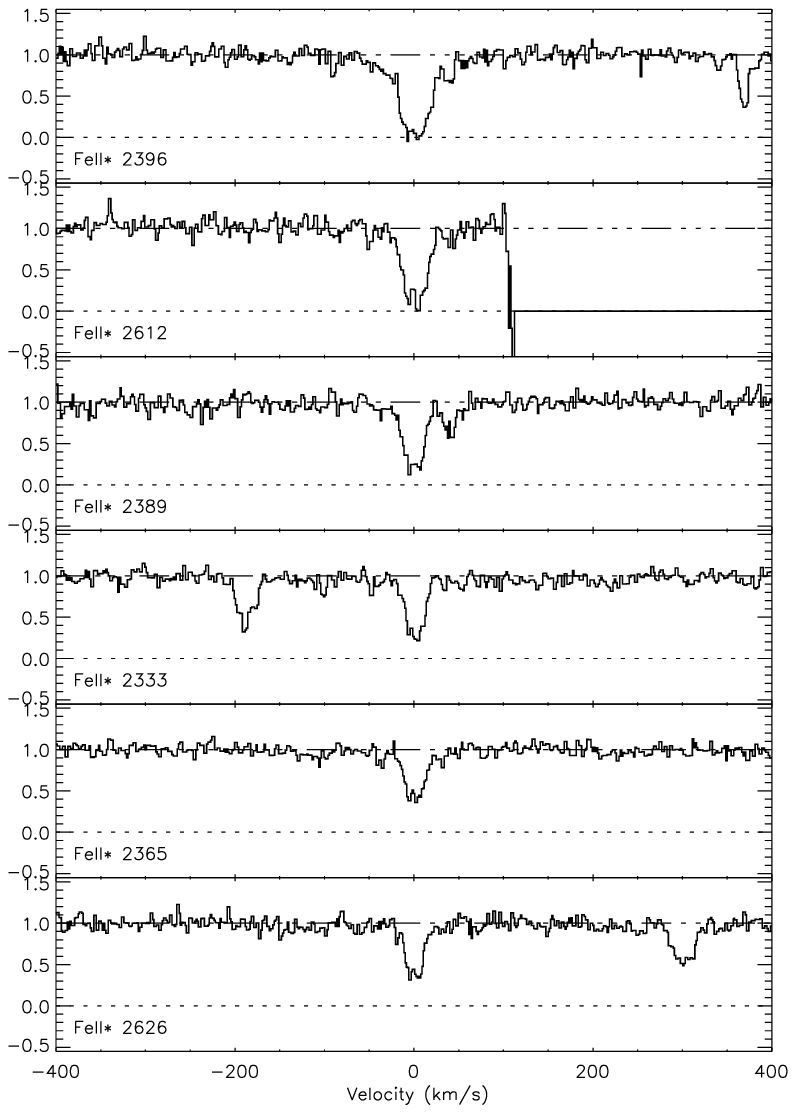,width=6.0in}}
\caption{Same as Figure~\ref{fig:MgIIspec} but for \ion{Fe}{2}* 
transitions. 
\label{fig:FeII*spec}}
\end{figure}

\begin{figure}
%\epsscale{0.9}
%\plotone{figbep2i.ps}
\centerline{\psfig{file=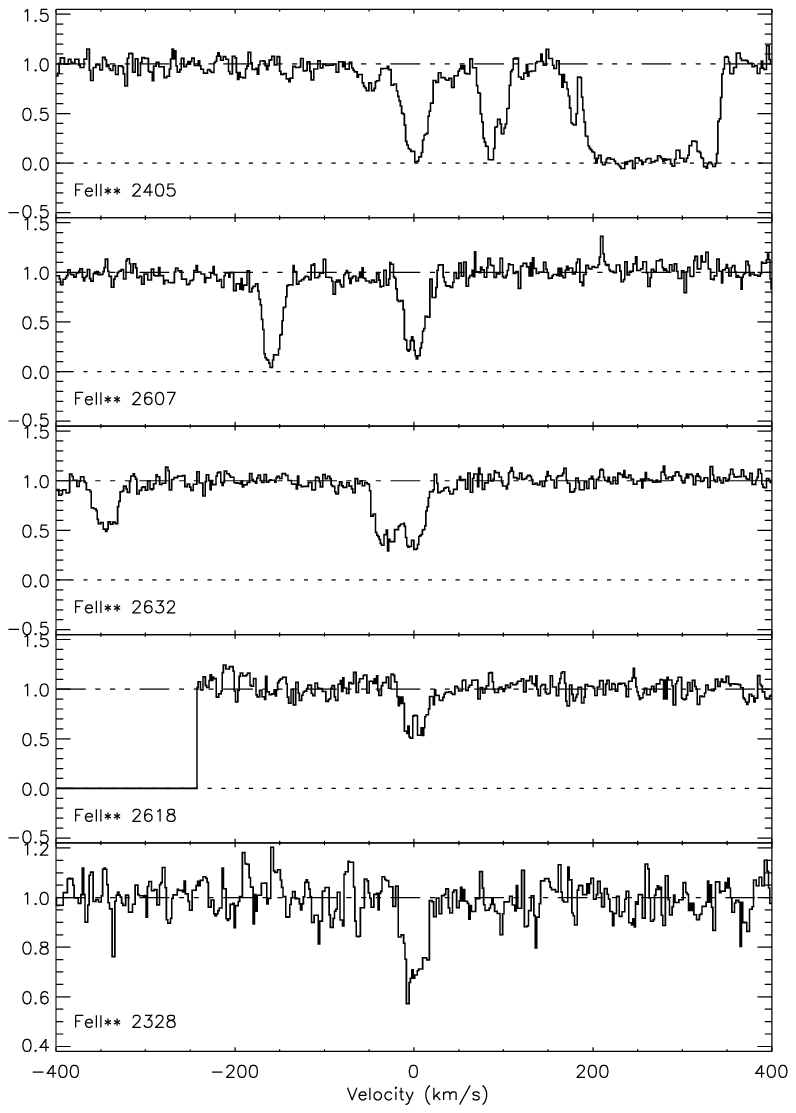,width=6.0in}}
\caption{Same as Figure~\ref{fig:MgIIspec} but for \ion{Fe}{2}** 
transitions. 
\label{fig:FeII**spec}}
\end{figure}

\begin{figure}
%\epsscale{0.9}
%\plotone{figbep2j.ps}
\centerline{\psfig{file=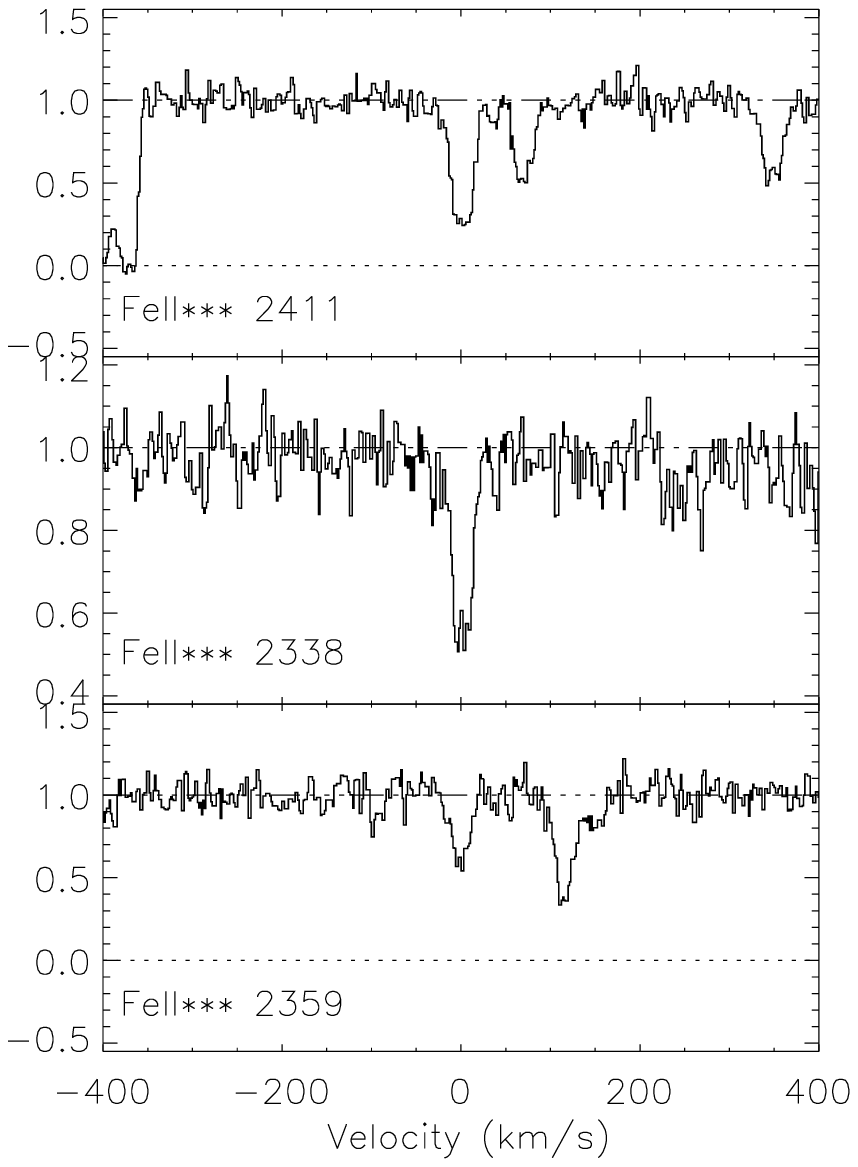,width=6.0in}}
\caption{Same as Figure~\ref{fig:MgIIspec} but for \ion{Fe}{2}*** 
transitions. 
\label{fig:FeII***spec}}
\end{figure}

\begin{figure}
%\epsscale{0.9}
%\plotone{figbep2k.ps}
\centerline{\psfig{file=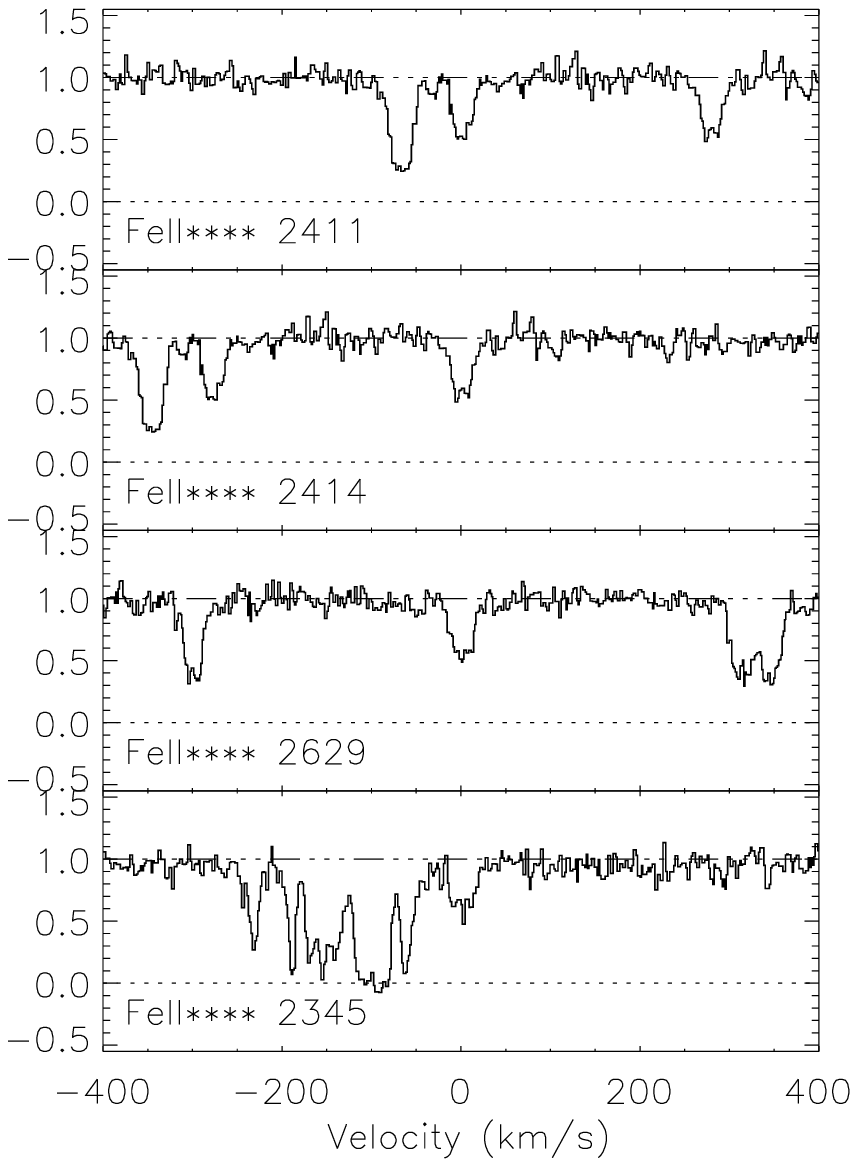,width=6.0in}}
\caption{Same as Figure~\ref{fig:MgIIspec} but for \ion{Fe}{2}**** 
transitions. 
\label{fig:FeII****spec}}
\end{figure}

\begin{figure}
%\epsscale{0.9}
%\plotone{figbep2l.ps}
\centerline{\psfig{file=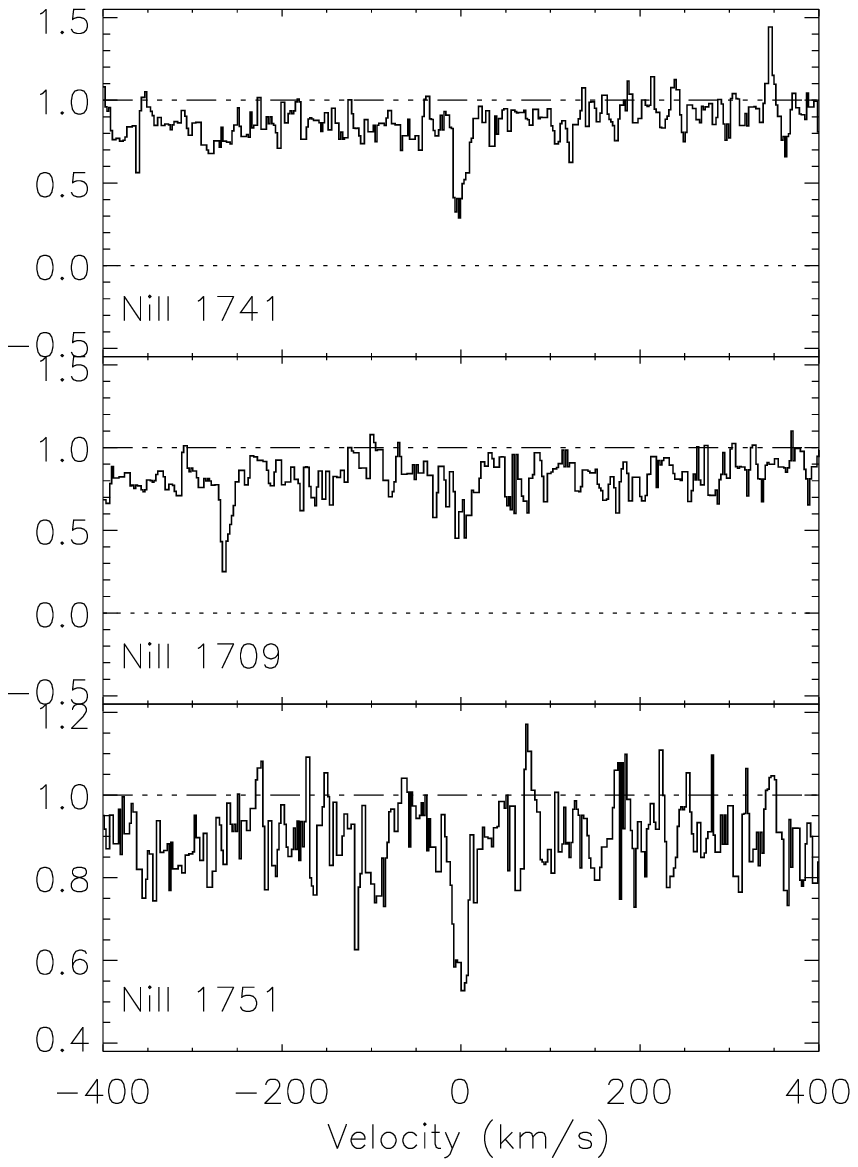,width=6.0in}}
\caption{Same as Figure~\ref{fig:MgIIspec} but for \ion{Ni}{2} 
transitions. 
\label{fig:NiIIspec}}
\end{figure}

\begin{figure}
%\epsscale{0.9}
%\plotone{figbep2m.ps}
\centerline{\psfig{file=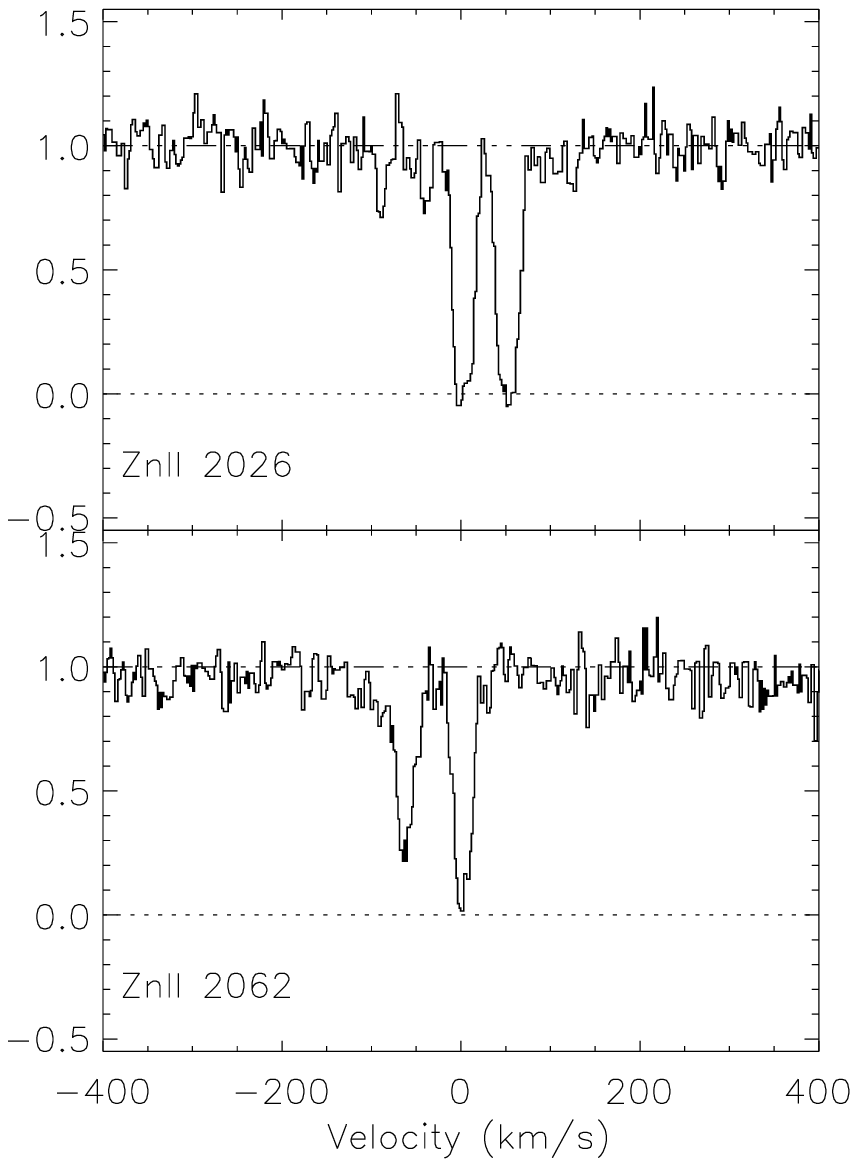,width=6.0in}}
\caption{Same as Figure~\ref{fig:MgIIspec} but for \ion{Zn}{2} 
transitions. 
\label{fig:ZnIIspec}}
\end{figure}

In Figures~\ref{fig:MgIIspec}--\ref{fig:ZnIIspec} we plot the
absorption profiles for all transitions of the various ionic species
as a function of velocity relative to the systemic redshift of
$z_1=1.54948$.  The majority of the lines exhibit a simple and
symmetric structure extending from about $-25$ to $+25$ km s$^{-1}$.
Lines of \ion{Fe}{2} and \ion{Mg}{2} exhibit in addition to the narrow
component absorbers in the range of $-150$ to $+50$ km s$^{-1}$ and
$-250$ to $+170$ km s$^{-1}$, respectively.  The absorption lines of
\ion{Al}{3} and \ion{Si}{2} exhibit an intermediate kinematic
structure with only a negative velocity extension.

The relatively low redshift precludes a direct detection of the
Ly$\alpha$ line within the wavelength range of our spectrum, but most
likely this system resembles a DLA in its neutral hydrogen column
density. In the discussion below we provide an estimate of the
approximate column density of neutral hydrogen by scaling the column
of weakly-depleted Zn along with the average metallicity of QSO-DLAs
at a similar redshift range.

The different line profiles can be interpreted to arise from physical
features in the host galaxy in the following way.  First, the
detection of positive velocity structure in \ion{Fe}{2} and
\ion{Mg}{2} suggests that the burst was located on the far side of
the host galaxy such that the disk and/or halo rotation are traced in
both negative and positive velocity.  The much larger extension at
negative velocity is expected since the burst is embedded in the disk
and does not trace its full extent on the far side.  The velocity
range of $|v|\lesssim 250$ km s$^{-1}$ is typical of the high velocity
and halo gas within our galaxy \citep{alb+93} and matches a range of
velocities for quasar absorption line systems known to probe galaxy
halos \citep{ell+03}.  It is possible that some of the higher
velocity absorption may arise from infall and/or outflow of metal
enriched gas.  However, this interpretation is difficult to reconcile
with the narrowness of the positive and negative velocity features
(Figures~\ref{fig:MgIIspec} and \ref{fig:FeIIspec}), which show line
widths of $b\approx 10$ km s$^{-1}$.

On the other hand, the main absorption component centered on zero
velocity is physically co-located with the burst, based on its unusual
properties as derived below, and the possibility that it is actually
influenced by the burst \citep{bpf+05}.  We show that this region is
unlike any that have been found in quasar spectra, but it is generally
similar to absorbers in other GRB spectra.  This suggests a fairly
compact region with a cross-section that is too small to be probed in
quasar sight lines, most likely a molecular cloud or an individual
star-forming region.

\section{Elemental Abundances within the Host Galaxy of GRB 051111}
\label{sec:abs}

\begin{figure}
%\epsscale{1}
%\plotone{cog.ps}
\centerline{\psfig{file=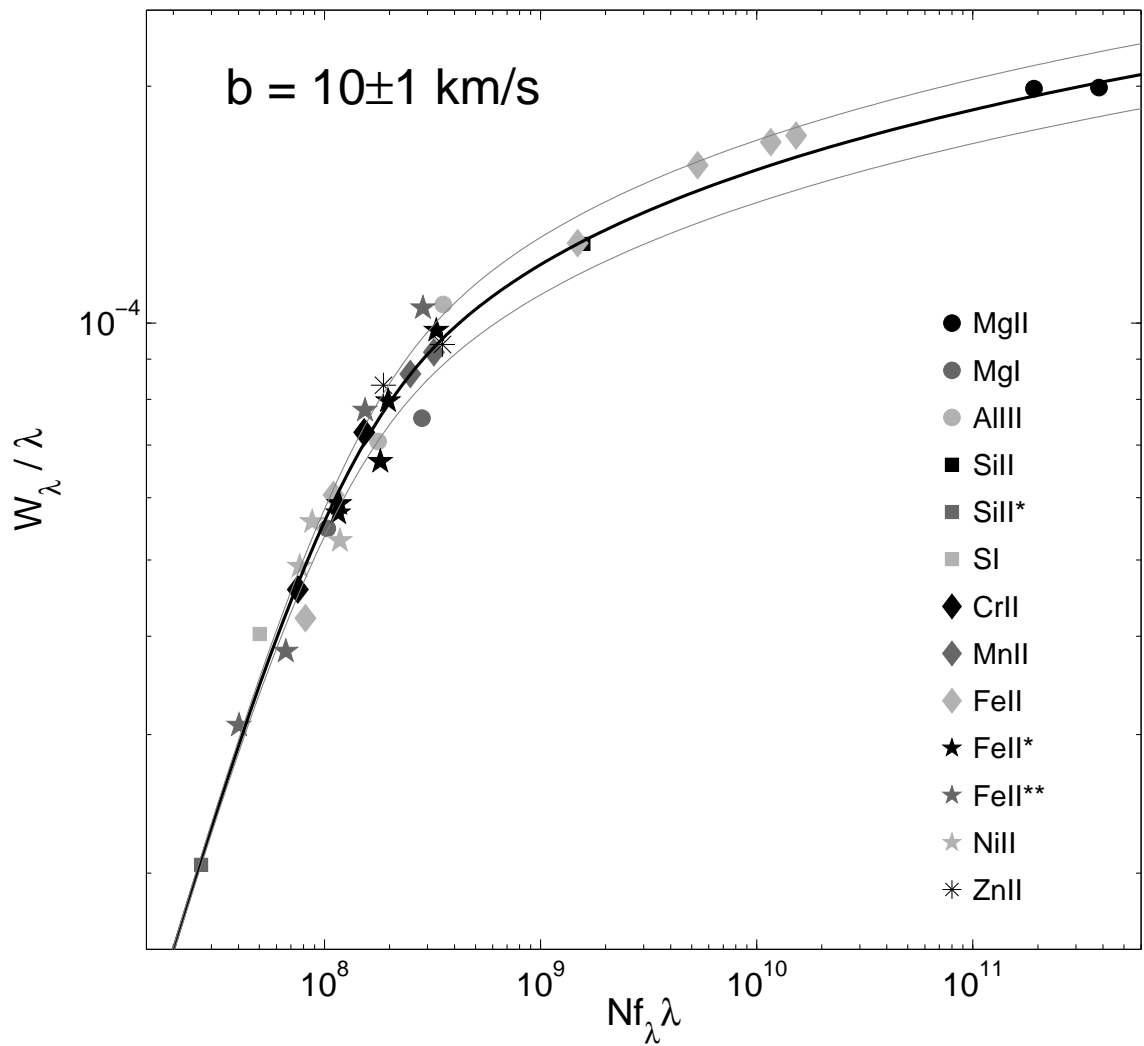,width=6.0in}}
\caption{Curve of growth (COG) for the host galaxy system of \grb.  We
constructed the COG by iteratively fitting for the column densities of
individual ions and for the Doppler parameter, $b$, which we assumed
to have a single value.  Transitions in the linear part of the COG
lead to well-determined columns.  Transitions on the flat portion of
the COG are sensitive to the value of $b$.
\label{fig:cog}} 
\end{figure}

We employ multiple techniques to investigate in detail the abundance
patterns of the narrow component centered at $v=0$ km s$^{-1}$, and
those of the extended positive and negative velocity structures.
Since some of the absorption features are saturated we make use of a
curve-of-growth (COG) analysis \citep{spi78}, which allows us to
correlate the equivalent widths
\begin{equation}
W_\lambda = \frac{2bF(\tau_0)\lambda}{c},
\end{equation}
with the line center's optical depth given by
\begin{equation}
\tau_0=\frac{\pi^{1/2}e^2f_\lambda\lambda N}{m_ecb}=1.496\times 
10^{-15} \frac{f_\lambda (\lambda/{\rm \AA})N}{(b/{\rm km\,s^{-1}})},
\label{eqn:tau}
\end{equation}
using the function 
\begin{equation}
F(\tau_0)=\int_{0}^{\infty}[1-{\rm exp}(-\tau_0{\rm e}^{-x^2})]
{\rm d}x.
\end{equation}
Here $f_\lambda$ is the oscillator strength, $W_\lambda$ is the
equivalent width, $\lambda$ is the rest wavelength, and we assume that
all species share the same value of the Doppler parameter, $b$.  We
fit iteratively for $b$ and the column density of each species (e.g.,
\citealt{sff03}).  The resulting best-fit COG, with a Doppler 
parameter, $b\approx 10$ km s$^{-1}$, is shown in
Figure~\ref{fig:cog}, and the column densities are listed in
Table~\ref{tab:columns}.

From the combined COG analysis we note that the best fit value of
$b=10$ km s$^{-1}$ is comparable to the observed values of \ion{Ca}{2}
line widths for atomic filaments and translucent molecular clouds in
our galaxy (e.g., \citealt{pen+00,pen+93}), as well as for the
well-studied sight lines of Zeta Oph, for which HST observations of
ions ranging from \ion{Mg}{2} to \ion{Zn}{2} show absorption over a
velocity range of 20 km s$^{-1}$, and structure with typical widths of
approximately $5-10$ km s$^{-1}$ \citep{ss96}.

To obtain additional information on the column density velocity
structure, $N(v)$, and the abundances of elements as a function of velocity,
we also use the apparent optical depth method \citep{ss91} where
\begin{equation}
N=\int N(v){\rm d}v = \frac{m_ec}{\pi e^2f\lambda}\int \tau(v)
{\rm d}v,
\end{equation}
and the optical depth
\begin{equation}
\tau(v)={\rm ln}[I_0(v)/I(v)].
\end{equation}
This method has the advantage that it makes no {\it a priori}
assumptions about the functional form of the velocity distribution,
and at the same time it incorporates information from a wide variety
of lines with different oscillator strengths.  By combining the
optical depths of these lines, and eliminating the contribution from
high optical depth and saturated velocity channels in the coadded
velocity spectrum, an improved estimate of both the column density and
the velocity structure of the line profile is obtained, which include
a measurement of the velocity structure within the cores of the lines.

\begin{figure}
%\epsscale{1}
%\plotone{ZnCr.ps}
\centerline{\psfig{file=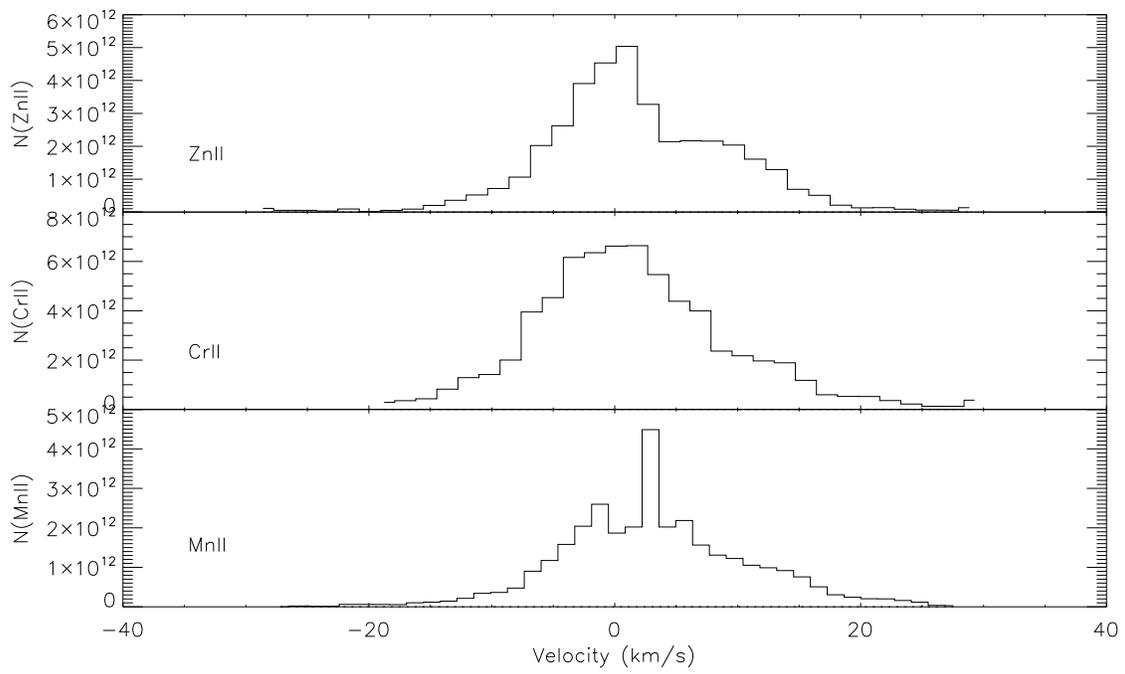,width=6.0in}}
\caption{Stacked plot of $N(v)$ of the ions \ion{Zn}{2},
\ion{Cr}{2}, and \ion{Mn}{2}, showing the strongly peaked, and 
narrow component at $-30<v<30$ km s$^{-1}$.
\label{fig:zncr}}
\end{figure}

\begin{figure}
%\epsscale{1}
%\plotone{AlSi.ps}
\centerline{\psfig{file=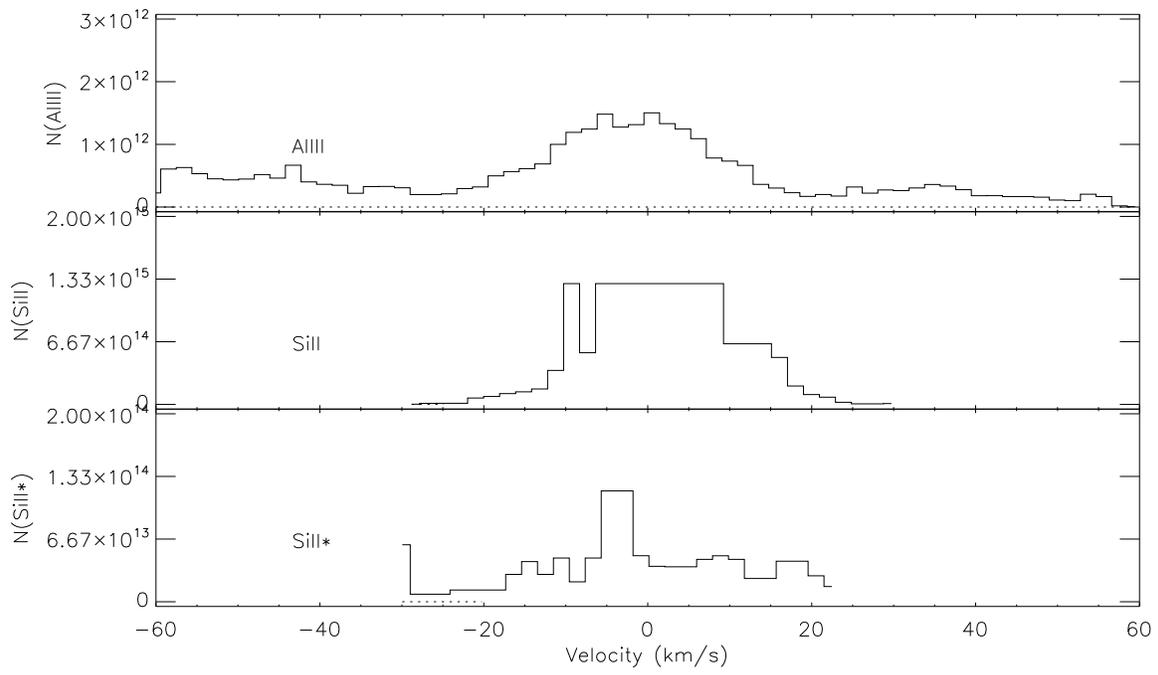,width=6.0in}}
\caption{Stacked plot of $N(v)$ for the \ion{Al}{3}, \ion{Si}{2} and 
\ion{Si}{2}* within \grb. The \ion{Si}{2} line is saturated, and our 
resulting column density is a lower limit.  The \ion{Si}{2}* column
density is clearly much weaker compared to the ground-state then is
observed in the case of \ion{Fe}{2}.
\label{fig:alsi}}
\end{figure}

We apply the apparent optical depth technique to all of the detected
absorbing species and summarize the resulting column densities in
Table~\ref{tab:columns}; the $N(v)$ profiles are shown in
Figures~\ref{fig:zncr}--\ref{fig:alsi}.  We also include in
Table~\ref{tab:columns} the adopted column density for each species
which is either derived from the apparent optical depth approach or an
average with the COG values, depending on the number of transitions,
and degree of saturation.  Based on the shapes of the various lines we
define three velocity ranges --- $z_{1a}$: $v<-30$ km s$^{-1}$,
$z_{1b}$: $-30<v<+30$ km s$^{-1}$, and $z_{1c}$: $v>+30$ km s$^{-1}$
--- and determine the column densities in each range.  This is
particularly useful for the \ion{Fe}{2} and \ion{Mg}{2} which exhibit
a wide velocity range.  We find that the dominant $z_{1b}$ system
accounts for about $85\%$ of the \ion{Fe}{2} column and $\gtrsim 70\%$
of the \ion{Mg}{2} column.

We estimate the neutral hydrogen column density using the column
density of \ion{Zn}{2} since Zn is a non-refractory iron peak element
and its gas-phase abundance should therefore closely match the gas
metallicity.  Based on a value of ${\rm log}\,N({\rm ZnII})=13.58\pm
0.15$ and a typical metallicity of ${\rm [Zn/H]}\approx -1$ for DLAs
at $z\sim z_1$ \citep{aep+05}, we derive ${\rm log}\,N({\rm
HI})\approx 21.9$, indicating a DLA with an \ion{H}{1} column density
that is similar to those of some other GRB absorption systems (e.g.,
\citealt{vel+04,bpc+05}), and is significantly larger than in typical
QSO-DLAs.  This is not surprising since the \ion{Zn}{2} column density
exceeds the highest column densities measured in QSO-DLAs by about 0.5
dex, and the median value by nearly 1.5 dex (Figure~\ref{fig:xzi}).
As we show below, the detailed abundance pattern of the main
absorption component ($z_{1b}$) is markedly different than in QSO-DLAs
and along with the large inferred value of $N({\rm HI})$ reflects the
fact that it is a region of the ISM that is generally missed in quasar
sight lines.

\section{Depletion Patterns and Dust Within the \grb\ Host Galaxy}
\label{sec:core}

\begin{figure}
%\epsscale{1}
%\plotone{x-zi.ps}
\centerline{\psfig{file=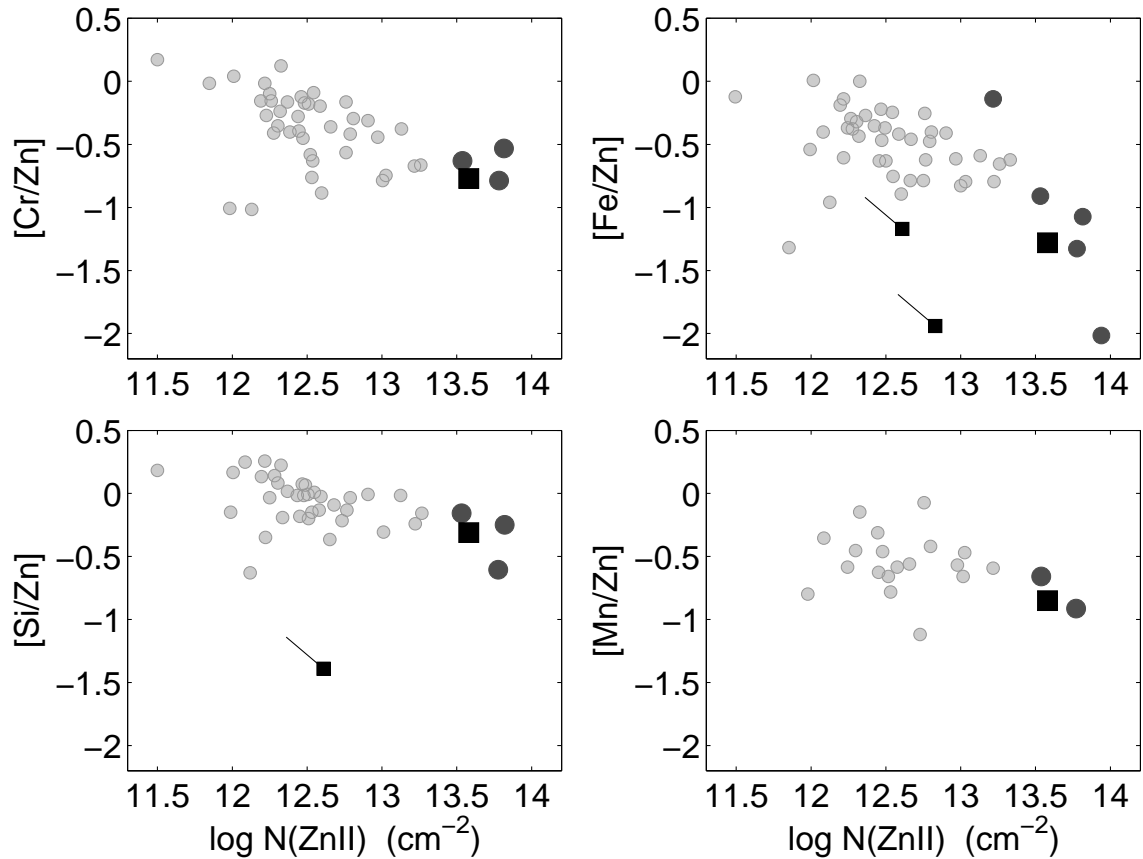,width=6.0in}}
\caption{Column density ratios from \grb of various species compared 
to the non-refractory \ion{Zn}{2} (squares), along with the values for
past GRB-DLAs and for QSO-DLAs (circles). The DLA sample is taken from
\citep{sf04}, and is consistent with a newer DLA sample from the Keck
HIRES spectrograph. Small squares show the range of values possible
for absorption from the system at larger velocities outside the main
-30 $<$ v $<$ 30 km s$^{-1}$, and where we have lower limits, the
ratios follow the vector indicated in the figure.
\label{fig:xzi}}
\end{figure}

\begin{figure}
%\epsscale{1}
%\plotone{newabundance.ps}
\centerline{\psfig{file=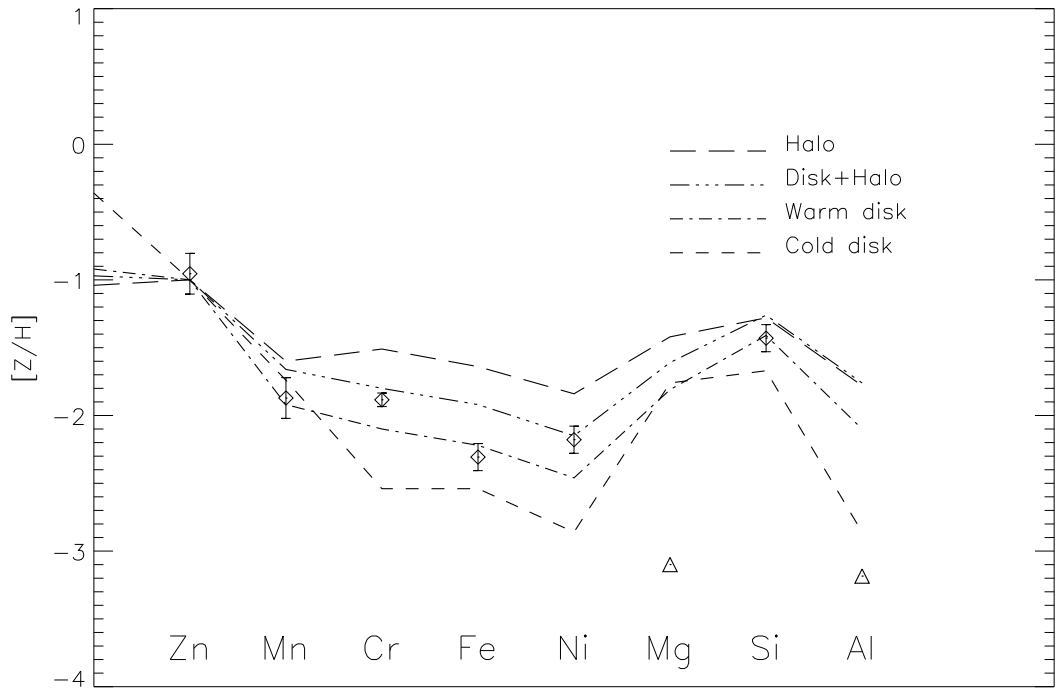,width=6.0in}}
\caption{Depletion pattern for the main absorber ($z_{1b}$) in the 
host galaxy of \grb.  The pattern of depletion is typical for warm
disk clouds such as those of observed in the Milky Way \citep{ss96}.
\label{fig:dep}}
\end{figure}

The difference in the velocity structure of various species, coupled
with the expected difference in depletion for the non-refractory
(e.g., Zn) and refractory (e.g., Fe) elements, indicates that the
sight line to \grb\ likely probes various components of the host
galaxy's interstellar medium.  We begin by investigating the main
absorption component $z_{1b}$.  The column density ratios of various
species compared to the non-refractory \ion{Zn}{2} are shown in
Figure~\ref{fig:dep}, along with the values for past GRB-DLAs and for
QSO-DLAs in Figure~\ref{fig:xzi}.

We find that for $z_{1b}$ the Cr to Zn ratio is ${\rm [Cr/Zn]}=-0.8\pm
0.2$, which is at the low end of the distribution for QSO-DLAs with
$\langle {\rm [Cr/Zn]}\rangle= -0.3\pm 0.3$ (Figure~\ref{fig:xzi}). 
Since Cr and Zn are produced in the same nucleosynthetic pathway,
differences in the abundances of these elements arise from
differential dust depletion.  Similarly, the ratio ${\rm
[Fe/Zn]}=-1.3\pm 0.2$ is significantly lower than in QSO-DLAs for
which $\langle {\rm [Fe/Zn]}\rangle=-0.5\pm 0.3$.  We find the same
results for the ratios of Si and Mn relative to Zn.  In all cases our
elemental abundances are referenced relative to the Solar values in
\citet{lod03}.  Taken in conjunction with the unusually large \ion{Zn}{2}
column density compared to QSO-DLAs, we conclude that the main
absorption component is dense and strongly dust-depleted.

The large \ion{Zn}{2} column density and strong depletion are
remarkably similar to those measured in past GRB absorption systems
\citep{sf04}.  This suggests that GRB sight lines probe similar 
regions at various redshifts reflecting a possible uniformity in the
environmental conditions that support GRB progenitor formation and
possibly star formation in general.  Similarly, the strong dust
depletion indicates that similar regions are missed in quasar sight
lines not just because of their small cross-section, but also because
of the associated dust extinction.

Given that Zn is largely undepleted, we estimate the dust content
along the sight line to \grb\ to be  about 2.5 times higher than in
typical DLA gas, but comparable, per unit metallicity, to Milky Way
gas.  Applying our adopted metallicity of $1/10$ solar we find that
the dust-to-gas ratio is approximately $1/12$ that of Milky Way gas,
scaling the relation described in \citet{pet03}.  With our estimated
column density of ${\rm log}\,N({\rm HI})\approx 21.9$, we then expect
about 0.55 magnitudes of $V$-band extinction toward \grb\, using the
relation $\langle N(HI)/Av\rangle=1.5\times 10^{21}$ cm$^{-2}$
mag$^{-1}$ \citep{ds94}. Using an SMC extinction curve appropriate for
low-metallicity gas \citep{blm+85}, we derive that the \grb\ sight
line has an extinction of $A(1500)\sim 2.5$ magnitudes in the rest
frame UV.

As described in detail in \citet{ss96}, depletion patterns can reveal
the presence of warm and cold gas along the line of sight.  The full
depletion pattern for the $z_{1b}$ component is shown in
Figure~\ref{fig:dep} in comparison to the four typical patterns
observed in the Milky Way of warm halo (WH), warm disk+halo (WDH),
warm disk (WD), and cool disk (CD) clouds \citep{ss96}.  We note that
the Mg absorption lines are heavily saturated and the inferred
abundance is therefore a lower limit.  Similarly, in the case of Al,
only the \ion{Al}{3} transition is observed, and we are therefore
unable to account for any additional \ion{Al}{2} gas.  Taking these
effects into account we find that the observed abundances closely
match the WD pattern, or marginally the WDH pattern.  The more weakly
depleted WH and the strongly-depleted CD do not provide an adequate
fit to the data.

So far we have discussed only the main component of the line
absorption profiles, but it is clear from the different shape of the
\ion{Fe}{2} and \ion{Mg}{2} lines compared to the other species that 
the depletion pattern changes across the line profile.  We find that
${\rm [Fe/Zn]}>-1.9$ ($z_{1a}$) and ${\rm [Fe/Zn]}>-1.2$ ($z_{1c}$),
along with the limits of ${\rm log}\,N({\rm ZnII})\lesssim 12.7$ are
similar to those found in QSO-DLAs (Figure~\ref{fig:xzi}), but are
different from the depleted $z_{1b}$ component.  This suggests that
the sight line to \grb\ probes in addition to the dense component in
the GRB local environment, a region in the halo of the host galaxy,
which is typical of the sight lines probed by background quasars.
Given the combination of kinematic structure and abundance patterns we
find that the GRB probably exploded away from the center of the
galaxy, most likely in the spiral arm of a highly inclined disk.

\section{Fine-Structure Excitation of \ion{Fe}{2} and \ion{Si}{2}}
\label{sec:exc}

In addition to the various transitions discussed above, from which we
deduce a warm and depleted environment in the $z_{1b}$ absorber
co-located with the GRB, we also detect absorption lines arising from
all of the available ground-level fine-structure states of \ion{Fe}{2}
and \ion{Si}{2}.  A detailed analysis of these lines and the inferred
physical conditions are summarized in a companion paper
\citep{bpf+05}, but we investigate this here in the context of the 
inferred depletion pattern.  

\begin{figure}
%\epsscale{1}
%\plotone{allFe.ps}
\centerline{\psfig{file=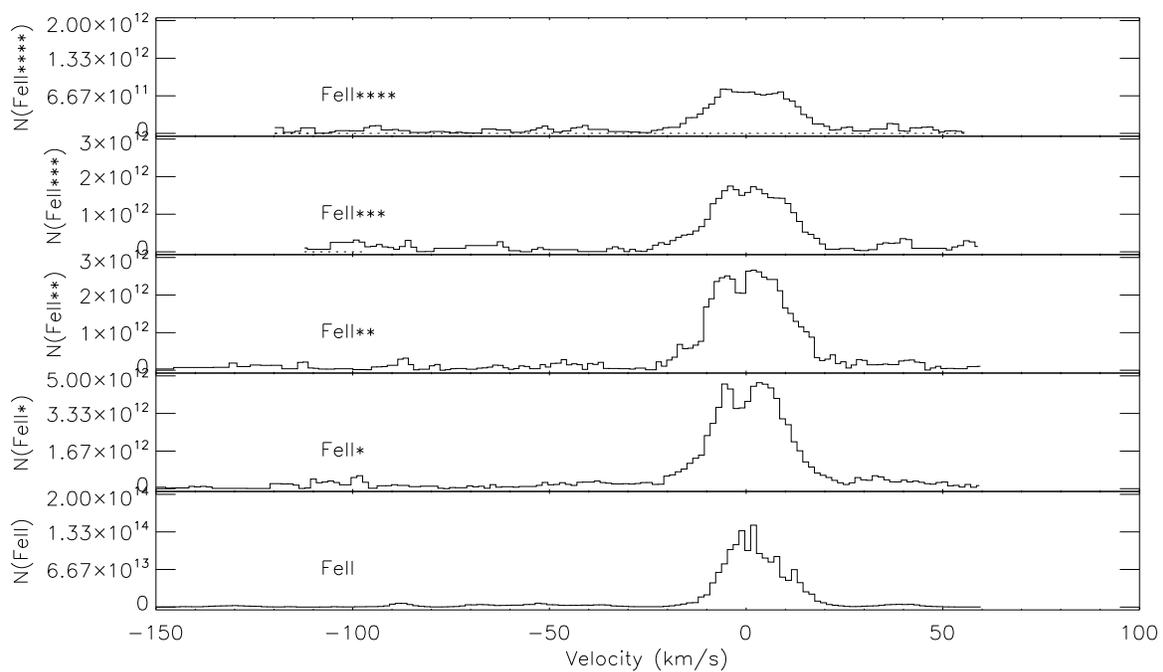,width=6.0in}}
\caption{Stacked plot of $N(v)$ for the fine-structure excited 
\ion{Fe}{2} levels within the main absorption system of \grb, 
plotted across the entire velocity range of detected \ion{Fe}{2}. The
kinematics of the \ion{Fe}{2} across the fine structure levels shows a
strong and relatively narrow pair of peaks with slightly broader
absorption in the higher excitation levels.
\label{fig:allfec}}
\end{figure}

As shown in Figure~\ref{fig:allfec}, the velocity structure of the
\ion{Fe}{2} fine-structure levels is similar to that of the dense 
and depleted component, $z_{1b}$.  The \ion{Si}{2}* transition is
weaker and as a result the velocity structure is not clear, but the
strongest absorption arises from $v\approx 0$ km s$^{-1}$
(Figure~\ref{fig:allfec}).  We use the apparent optical depth method
over the full velocity range of $|v|<30$ km s$^{-1}$ to derive the
column densities of the fine-structure states (see
Table~\ref{tab:columns}).  We find that ${\rm log}\,N({\rm
SiII^*})=14.96$ is somewhat higher than the column densities measured
in GRB\,050505 (${\rm log}\,N\approx 14.7$; \citealt{bpc+05}),
GRB\,030323 (${\rm log}\,N\approx 14.2$; \citealt{vel+04}), and
GRB\,020813 (${\rm log}\,N\approx 14.3$; \citealt{sf04}).

The excitation of the fine-structure levels requires either a
combination of a high temperature and a high gas volume density, or an
intense IR or UV radiation field.  As we show in \citet{bpf+05}, the
radiation field may be due to the GRB itself, but in the case of
excitation by the ambient radiation field due to star formation
activity, the inferred luminosity and size of the region would
naturally lead to a warm, dust-depleted environment.

\begin{figure}
%\epsscale{1}
%\plotone{chisq.contour.new.ps}
\centerline{\psfig{file=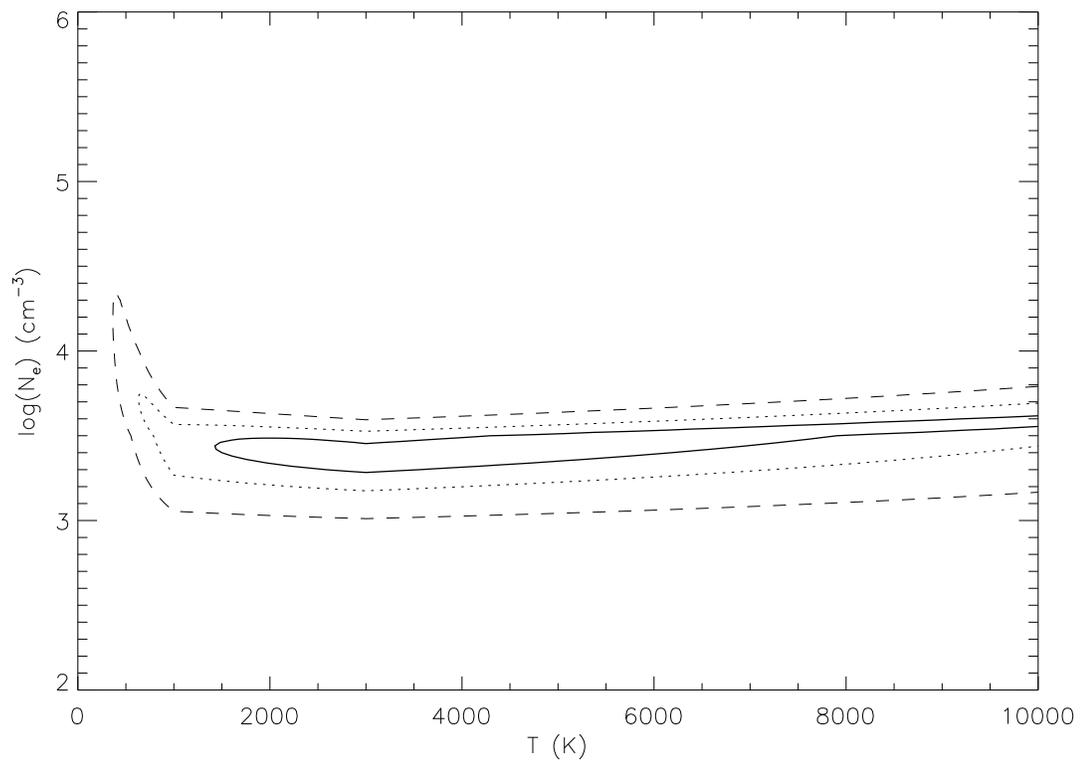,width=6.0in}}
\caption{Reduced $\chi^{2}$ contours for the \ion{Fe}{2} 
fine-structure column densities compared to the ground-state compared
with predictions from \citet{khb+88}. The contours show values of
$\chi^2_r=1$ (solid), 2 (dotted) and 5 (dashed), giving a range of
possible density and temperature within the \grb\ absorber, assuming
collisional excitation.  
\label{fig:chisq}}
\end{figure}

In particular, excitation of the fine-structure levels by an IR
radiation field, along with the observed ratios of the \ion{Fe}{2} and
\ion{Si}{2} fine-structure levels, are indicative of a warm dust 
spectrum, $F_\nu\propto\nu^{2.2}$ with $T\gtrsim 600$ K.  The
alternative interpretation of collisional excitation similarly leads
to a large electron volume density, $n_e\gtrsim 10^3$ cm$^{-3}$, and
temperature, $T_e\sim 10^3$ K (Figure~\ref{fig:chisq}). In both
scenarios we expect that the depletion pattern would be similar to
that of a warm disk environment, and with a significant depletion.

\section{GRB Host Galaxy Imaging}
\label{sec:host}

\begin{figure}
%\epsscale{1}
%\plotone{image.ps}
\centerline{\psfig{file=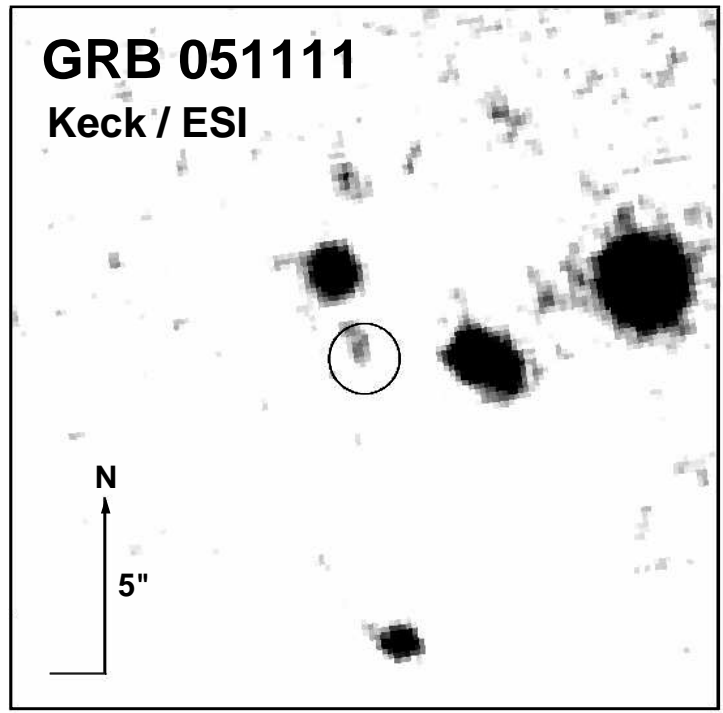,width=6.0in,angle=270}}
\caption{Keck ESI image of the field near \grb, showing the error 
circle of the coordinates for the \grb (shown with the circle), and
two adjacent galaxies probably responsible for the intervening
\ion{Mg}{2} and \ion{Fe}{2} absorption lines seen at lower redshifts
of $z_2=1.18975$, $z_{3a}=0.82761$ and $z_{3b}=0.82698$.  The $2-3''$
offset from the centers of these adjacent galaxies amounts to a
distance from the center of the galaxy of $15-20$ kpc. 
\label{fig:image}}
\end{figure}

In an attempt to place the information from the absorption spectrum of
\grb\ in the overall context of the host galaxy properties, we 
observed the position of the burst with the Echellete Spectrograph and
Imager (ESI) mounted on the Keck II telescope on 2005 November 30.  A
total of 1500 s were obtained in $R$-band.  At the position of the
afterglow we detect a faint extended source which we identify as the
host galaxy (Figure~\ref{fig:image}).  A comparison to two nearby and
unsaturated stars in the USNO-B catalog indicates a brightness of
$R=26\pm 0.3$ mag for this object.  At the redshift $z_1$ and assuming
a spectrum $F_\nu\propto \nu^{-1}$, this translates to an absolute
rest-frame $B$-band magnitude of about $-18.6$ or $L\approx 0.1$ L*.
This value is similar to that of other GRB host galaxies which at a
similar redshift range from about $-17$ to $-21$ mag.  

Since the observed $R$-band probes the rest-frame UV emission from the
host galaxy, we can roughly estimate the star formation rate in the
host galaxy.  Using the relation of \citet{ken98}, ${\rm
SFR}=1.4\times 10^{-28} L_\nu$, we find that ${\rm SFR}\approx 3$
M$_\odot$ yr$^{-1}$.  Clearly, this value is subject to a large upward
correction due to possible dust extinction.  If the estimated
extinction provided above based on the depletion pattern in the GRB
local environment is representative, then it is possible that the star
formation rate easily exceeds $30$ M$_\odot$ yr$^{-1}$.

\section{Intervening Systems}
\label{sec:inter}

Within the spectrum of \grb\ we detect intervening absorption systems
at $z_2=1.18975$ from absorption by \ion{Mg}{2}, \ion{Mg}{1}, and
\ion{Fe}{2}. The column densities for these systems derived from
optical depth and COG fitting of the absorption lines are ${\rm
log}\,N({\rm MgII})=14.33\pm 0.05$, ${\rm log}\,N({\rm MgI}) =12.58\pm
0.1$, and ${\rm log}\,N({\rm FeII})=14.31\pm 0.05$.  An additional
pair of absorption systems are seen at $z_{3a}=0.82761$ and
$z_{3b}=0.82698$ in the species \ion{Mg}{1}, \ion{Mg}{2}, and
\ion{Fe}{2}. For the absorption system at $z_{3a}=0.82761$, we
estimate column densities of ${\rm log}\,N({\rm MgII})=13.28\pm 0.1$,
${\rm log}\,N({\rm MgI})=12.38\pm 0.1$, and ${\rm log}\,N({\rm
FeII})=12.95\pm 0.05$. The column densities of the absorption lines at
$z_{3b}=0.82698$ are ${\rm log}\,N({\rm MgII})=13.1\pm 0.15$, and
${\rm log}\,N({\rm FeII})=12.70\pm 0.05$.

\citet{bbp95} conducted a survey of \ion{Mg}{2} absorption from
the disks and halos of 17 low-redshift galaxies, and studied the
correlation between the impact parameter, $\rho$, and the line
equivalent width.  The sample includes a range of $\rho\approx 2-113$
kpc, and equivalent widths from 0.05 to 2 \AA.  Systems with
\ion{Mg}{2} equivalent widths $>0.1$\AA\ appear to have smaller
impact parameters, $\rho<20-30$ kpc, while the largest impact
parameter systems are seen to have much smaller equivalent widths.
The $z_2$ and $z_3$ intervening systems in the spectrum of \grb\
exhibit equivalent widths of 1.7\AA\ and 0.2\AA, respectively.  Thus,
both of these intervening systems are at the high end of the
\citet{bbp95} sample, indicating small impact parameters.  

Our image of the field of \grb\ reveals two nearby galaxies, which are
significantly brighter than the GRB host, one located $2''$ to the
north and the other about $2.5''$ to the west
(Figure~\ref{fig:image}).  While we do not have redshifts for these
galaxies, it is likely that they are responsible for the intervening
systems $z_2$ and $z_{3ab}$.  The derived impact parameters relative
to the GRB position are about 17 and 20 kpc, indicating that the
intervening absorbers are similar to those found by \citet{bbp95}.

\section{Discussion and Conclusions}
\label{sec:disc}

The high resolution spectrum of \grb\ reveals a range of interstellar
abundance patterns within the host galaxy.  In addition to warm halo
gas which is typical of quasar sight lines and which represents the
bulk of the galaxy cross-section, we detect a high column density,
kinematically cold region of dust-depleted gas, which is typical of a
warm disk abundance pattern.  The observed abundance pattern in this
component is similar to those in several other GRB absorption spectra
obtained in the past.  The weaker Fe and Mg absorption with a wider
positive and negative velocity extension of several hundred km
s$^{-1}$ suggests that the GRB location was offset from the center of
the galaxy.  While the Ly$\alpha$ line is not detected in our
spectrum, a reasonable value of the metallicity $Z\sim 0.1$ Z$_\odot$,
combined with a large \ion{Zn}{2} abundance indicates that the host is
likely a DLA with ${\rm log}N\,({\rm HI})\approx 21.9$, higher than
typical QSO-DLAs but in good agreement with other GRB-DLA systems
\citep{vel+04,bpc+05,cpb+05}.

The detection of strong absorption from \ion{Fe}{2} and \ion{Si}{2}
fine-structure levels (discussed in detail in \citealt{bpf+05})
requires physical conditions that are in good agreement with the
conclusion that the local environment is warm and dust-depleted.  In
the case of collisional excitation the inferred temperature is about
$10^3$ K, while in the case of IR pumping dust re-processing is
probably essential for generating the required IR radiation field.

We finally identify the host galaxy and the likely counterparts of the
two intervening systems detected in the spectrum.  As in the case of
previous GRB hosts, the galaxy is relatively faint, $L\sim 0.1$ L*.  A
comparison of the \ion{Mg}{2} column densities of the intervening
systems with the offsets of the two nearby galaxies relative to the
GRB position is in good agreement with previous correlations with
impact parameters found in the context of quasar studies.  If the
intervening systems occurred at a higher redshift in which we could
determine directly whether they are DLAs, one can imagine that similar
imaging, as well as follow-up spectroscopy, would provide direct
insight into the nature of DLAs, a question that remains difficult to
address in the context of QSO-DLAs.

\acknowledgements 
BEP would like to thank Pomona College for support from a Downing
Exchange fellowship to Cambridge University, and would like to thank
Max Pettini, Edward Jenkins, and Bruce Draine for helpful discussions.
EB is supported is supported by NASA through Hubble Fellowship grant
HST-01171.01 awarded by the Space Telescope Science Institute, which
is operated by AURA, Inc., for NASA under contract NAS 5-26555.

%\bibliographystyle{apj}
%\bibliography{journals_apj,refs}

\clearpage
\begin{deluxetable}{llllll}
\tablecolumns{6}
\tabcolsep0.2in\footnotesize
\tablewidth{0pc}
\tablecaption{Line Identification
\label{tab:lines}}
\tablehead {
\colhead {$\lambda_{\rm obs}$}         &
\colhead {Line}                        &
\colhead {$f_{ij}$}                    &
\colhead {$z$}                         &
\colhead {$W_0$}                       &
\colhead {${\rm log}\,N$}                \\
\colhead {(\AA)}                       &
\colhead {(\AA)}                       &
\colhead {}                            &
\colhead {}                            &
\colhead {(\AA)}                       &
\colhead {(cm$^{-2}$)}                       
}
\startdata
4353.26 &  \ion{Fe}{2}   2382.7650 & 0.32000 & 0.82698 & 0.09582 & 12.58 \\
4354.79 &  \ion{Fe}{2}   2382.7650 & 0.32000 & 0.82762 & 0.13127 & 12.78 \\
4358.60 &  \ion{Ni}{2}   1709.6042 & 0.03240 & 1.54948 & 0.09561 & 13.76 \\
4440.05 &  \ion{Ni}{2}   1741.5531 & 0.04270 & 1.54948 & 0.09213 & 13.64 \\
4455.97 &  \ion{Mg}{1}   1747.7937 & 0.00934 & 1.54948 & 0.09585 & 14.36 \\
4466.47 &  \ion{Ni}{2}   1751.9157 & 0.02770 & 1.54948 & 0.08595 & 13.75 \\
4607.70 &  \ion{S}{1}    1807.3113 & 0.11050 & 1.54948 & 0.07275 & 13.09 \\
4609.49 &  \ion{Si}{2}   1808.0130 & 0.00219 & 1.54948 & 0.22796 & 15.77 \\
4632.22 &  \ion{Si}{2}*  1816.9285 & 0.00166 & 1.54948 & 0.03723 & 14.55 \\
4660.28 &  \ion{Mg}{1}   1827.9351 & 0.02450 & 1.54948 & 0.13838 & 14.29 \\
4707.90 &  \ion{Mn}{2}   2576.8770 & 0.35080 & 0.82698 & 0.01619 & 11.72 \\
4709.55 &  \ion{Mn}{2}   2576.8770 & 0.35080 & 0.82762 & 0.03428 & 12.02 \\
4728.56 &  \ion{Al}{3}   1854.7164 & 0.53900 & 1.54948 & 0.19586 & 12.97 \\
4741.76 &  \ion{Mn}{2}   2594.4990 & 0.27100 & 0.82762 & 0.02474 & 11.98 \\
4749.14 &  \ion{Al}{3}   1862.7895 & 0.26800 & 1.54948 & 0.13177 & 12.96 \\
4750.46 &  \ion{Fe}{2}   2600.1730 & 0.22390 & 0.82698 & 0.04929 & 12.37 \\
4752.13 &  \ion{Fe}{2}   2600.1730 & 0.22390 & 0.82762 & 0.10517 & 12.76 \\
5108.88 &  \ion{Mg}{2}   2796.3520 & 0.61230 & 0.82698 & 0.14426 & 12.43 \\
5110.67 &  \ion{Mg}{2}   2796.3520 & 0.61230 & 0.82762 & 0.22484 & 12.83 \\
5122.00 &  \ion{Mg}{2}   2803.5310 & 0.30540 & 0.82698 & 0.11927 & 12.64 \\
5123.79 &  \ion{Mg}{2}   2803.5310 & 0.30540 & 0.82762 & 0.19326 & 13.07 \\
5133.24 &  \ion{Fe}{2}   2344.2140 & 0.11400 & 1.18975 & 0.71346 & 14.09 \\
5165.59 &  \ion{Zn}{2}   2026.1360 & 0.48900 & 1.54948 & 0.19034 & 13.19 \\
5214.13 &  \ion{Mg}{1}   2852.9640 & 1.83000 & 0.82762 & 0.19037 & 12.12 \\
5217.66 &  \ion{Fe}{2}   2382.7650 & 0.32000 & 1.18975 & 1.11235 & 13.99 \\
5242.38 &  \ion{Cr}{2}   2056.2539 & 0.10500 & 1.54948 & 0.14935 & 13.46 \\
5258.72 &  \ion{Zn}{2}   2062.6640 & 0.25600 & 1.54948 & 0.17199 & 13.24 \\
5267.64 &  \ion{Cr}{2}   2066.1610 & 0.05150 & 1.54948 & 0.09474 & 13.42 \\
5693.73 &  \ion{Fe}{2}   2600.1730 & 0.22390 & 1.18975 & 1.23064 & 14.01 \\
5736.02 &  \ion{Fe}{2}   2249.8768 & 0.00182 & 1.54948 & 0.09490 & 14.80 \\
5763.81 &  \ion{Fe}{2}   2260.7805 & 0.00244 & 1.54948 & 0.13672 & 14.87 \\
5935.47 &  \ion{Fe}{2}** 2328.1112 & 0.03450 & 1.54948 & 0.07175 & 13.30 \\
5949.25 &  \ion{Fe}{2}*  2333.5156 & 0.07780 & 1.54948 & 0.15565 & 13.41 \\
5962.53 & \ion{Fe}{2}*** 2338.7248 & 0.08970 & 1.54948 & 0.10608 & 13.08 \\ 
5976.53 &  \ion{Fe}{2}   2344.2140 & 0.11400 & 1.54948 & 0.37211 & 14.05 \\
5978.53 & \ion{Fe}{2}**** 2345.0011 & 0.15300 & 1.54948 & 0.10861 & 12.84 \\
6016.33 & \ion{Fe}{2}*** 2359.8278 & 0.06790 & 1.54948 & 0.07552 & 13.03 \\
6030.93 &  \ion{Fe}{2}*  2365.5518 & 0.04950 & 1.54948 & 0.13941 & 13.49 \\
6053.64 &  \ion{Fe}{2}   2374.4612 & 0.03130 & 1.54948 & 0.29999 & 14.43 \\
6074.81 &  \ion{Fe}{2}   2382.7650 & 0.32000 & 1.54948 & 0.41259 & 13.59 \\
6091.62 &  \ion{Fe}{2}*  2389.3582 & 0.08250 & 1.54948 & 0.19033 & 13.50 \\
6109.46 & \ion{Fe}{2}*   2396.3559 & 0.31619 & 1.54948 & 0.28800 & 13.37 \\
6123.31 &  \ion{Mg}{2}   2796.3520 & 0.61230 & 1.18975 & 1.99290 & 13.93 \\
6131.92 & \ion{Fe}{2}*** 2405.1638 & 0.02600 & 1.54948 & 0.05041 & 13.22 \\
6133.08 &  \ion{Fe}{2}** 2405.6186 & 0.23700 & 1.54948 & 0.25156 & 13.31 \\
6139.03 &  \ion{Mg}{2}   2803.5310 & 0.30540 & 1.18975 & 1.68043 & 14.13 \\
6147.44 & \ion{Fe}{2}*** 2411.2533 & 0.21000 & 1.54948 & 0.18791 & 13.04 \\
6148.84 & \ion{Fe}{2}**** 2411.8023 & 0.21000 & 1.54948 & 0.10497 & 12.68 \\ 
6154.56 & \ion{Fe}{2}**** 2414.0450 & 0.17500 & 1.54948 & 0.10723 & 12.77 \\
6247.28 &  \ion{Mg}{1}   2852.9640 & 1.83000 & 1.18975 & 0.38539 & 12.24 \\
6569.70 &  \ion{Mn}{2}   2576.8770 & 0.35080 & 1.54948 & 0.23663 & 13.08 \\
6594.61 & \ion{Fe}{2}   2586.6500 & 0.36933 & 1.54948 & 0.06840 & 14.09 \\
6614.62 &  \ion{Mn}{2}   2594.4990 & 0.27100 & 1.54948 & 0.22355 & 13.23 \\
6629.09 &  \ion{Fe}{2}   2600.1730 & 0.22390 & 1.54948 & 0.44168 & 13.75 \\
6648.70 &  \ion{Fe}{2}** 2607.8664 & 0.11800 & 1.54948 & 0.20213 & 13.28 \\
6660.91 &  \ion{Fe}{2}*  2612.6542 & 0.12600 & 1.54948 & 0.25575 & 13.50 \\
6675.56 &  \ion{Fe}{2}** 2618.3991 & 0.05050 & 1.54948 & 0.10022 & 13.22 \\
6685.89 & \ion{Fe}{2}**** 2622.4518 & 0.05600 & 1.54948 & 0.04794 & 12.79 \\
6696.08 &  \ion{Fe}{2}*  2626.4511 & 0.04410 & 1.54948 & 0.15064 & 13.50 \\
6702.78 & \ion{Fe}{2}**** 2629.0777 & 0.17300 & 1.54948 & 0.12269 & 12.76 \\ 
6710.51 & \ion{Fe}{2}**  2632.1081 & 0.22860 & 1.54948 & 0.08600 & 13.40 \\
7129.24 &  \ion{Mg}{2}   2796.3520 & 0.61230 & 1.54948 & 0.55706 & 13.41 \\
7147.55 &  \ion{Mg}{2}   2803.5310 & 0.30540 & 1.54948 & 0.55682 & 13.70 \\

\enddata
\tablecomments{Absorption features identified in the spectrum of \grb.  
The columns are (left to right): (i) Observed wavelength, (ii) line
identification, (iii) oscillator strength, (iv) redshift of the line,
(v) rest-frame equivalent width, and (vi) logarithm of the column
density assuming the optically-thin case; in most cases this is a
lower limit since the lines are genearlly saturated.}
\end{deluxetable}

\clearpage
\begin{deluxetable}{lrrrrrrlll}
\tablecolumns{10}
\tablewidth{0pc}
\tablecaption{Column Densities of Ions in $z_1$
\label{tab:columns}}
\tablehead {
\colhead {}                 &
\multicolumn{3}{c}{COG}     &
\multicolumn{3}{c}{AOD}     &
\multicolumn{3}{c}{Adopted} \\\cline{2-4}\cline{5-7}\cline{8-10}		    
\colhead {Ion}		    &
\colhead {$z_{1a}$}         &
\colhead {$z_{1b}$}         &
\colhead {$z_{1c}$}         &
\colhead {$z_{1a}$}         &
\colhead {$z_{1b}$}         &
\colhead {$z_{1c}$}         &
\colhead {$z_{1a}$}         &
\colhead {$z_{1b}$}         &
\colhead {$z_{1c}$}         
}
\startdata
\ion{Mg}{2}     & 15.05    & $>14.78$ & 13.59    & 14.22    & $>13.88$ & 13.62    & $14.22\pm 0.08$ & $>14.70$        & $13.62\pm 0.08$  \\ 
\ion{Mg}{1}     & $<13.6$  & 14.85    & $<13.4$  & $<13.26$ & 14.72    & $<13.1$  & $<13.3$         & $14.72\pm 0.01$ & $<13.1$	      \\
\ion{Al}{3}     & $<13.2$  & 13.39    & $<12.4$  & 13.05    & 13.34    & $<12.4$  & $13.05\pm 0.20$ & $13.34\pm 0.10$ & $<12.4$	      \\
\ion{Si}{2}     & \nod     & \nod     & \nod     & 14.13    & 16.18    & $<14.5$  & $14.13\pm 0.10$ & $16.18\pm 0.20$ & $<14.5$	      \\
\ion{Si}{2}*    & \nod     & \nod     & \nod     & $<14.1$  & 14.96    & $<14.5$  & $<14.1$         & $14.96\pm 0.02$ & $<14.5$	      \\
\ion{S}{1}      & \nod     & \nod     & \nod     & $<12.3$  & 13.50    & $<12.6$  & $<12.3$         & $13.50\pm 0.02$ & $<12.6$	      \\
\ion{Cr}{2}     & $<13.2$  & 13.80    & $<11.8$  & $<12.9$  & 13.85    & $<11.9$  & $<12.9$         & $13.85\pm 0.10$ & $<11.9$	      \\
\ion{Mn}{2}     & $<10.4$  & 13.84    & $<11.5$  & $<10.4$  & 13.52    & $<11.6$  & $<10.4$         & $13.52\pm 0.15$ & $<11.6$	      \\
\ion{Fe}{2}     & 14.66    & 15.32    & 14.51    & 14.29    & 15.15    & 13.74    & $14.29\pm 0.10$ & $15.15\pm 0.10$ & $13.74\pm 0.10$ \\
\ion{Fe}{2}*    & $<12.5$  & 13.91    & \nod     & $<12.2$  & 13.90    & $<12.3$  & $<12.2$         & $13.90\pm 0.10$ & $<12.3$ \\
\ion{Fe}{2}**   & $<12.5$  & 13.70    & $<12.2$  & $<12.2$  & 13.70    & $<12.0$  & $<12.2$         & $13.70\pm 0.08$ & $<12.0$ \\
\ion{Fe}{2}***  & $<12.5$  & 13.54    & $<12.1$  & $<12.2$  & 13.52    & $<12.1$  & $<12.2$         & $13.53\pm 0.10$ & $<12.1$ \\
\ion{Fe}{2}**** & $<12.8$  & 13.21    & $<12.0$  & $<12.5$  & 13.15    & $<12.0$  & $<12.5$         & $13.18\pm 0.10$ & $<12.1$ \\
\ion{Ni}{2}     & $<13.5$  & 15.05    & $<12.9$  & $<14.2$  & 13.99    & $<12.8$  & $<13.5$         & $13.99\pm 0.35$ & $<12.8$	      \\
\ion{Zn}{2}     & $<11.7$  & 13.87    & $<11.9$  & $<11.8$  & 13.58    & $<12.0$  & $<11.8$         & $13.58\pm 0.15$ & $<12.0$        \\ 
\enddata
\tablecomments{Ionic column densities and abundances as derived 
from the curve-of-growth analysis and apparent optical depth 
technique.  The three velocity ranges are defined in 
\S\ref{sec:abs}.}
\end{deluxetable}

\clearpage

\clearpage

\clearpage

\clearpage

\clearpage

\clearpage

\clearpage

\clearpage

\clearpage

\clearpage

\clearpage

\clearpage

\clearpage

\clearpage

\clearpage

\clearpage

\clearpage

\clearpage

\clearpage

\clearpage

\clearpage

\clearpage

\end{document}